\documentclass[12pt]{article}

\usepackage{amssymb,amsmath,amsfonts,amssymb}
\usepackage{graphics,graphicx,color, subcaption}
\usepackage{empheq}

\def\@abssec#1{\vspace{.05in}\footnotesize \parindent .2in
{\bf #1. }\ignorespaces}

\graphicspath{{/EPSF/}{Figures/}}
\DeclareGraphicsExtensions{.eps}

\usepackage[margin=1.0in]{geometry}
\usepackage{tikz-cd}

\def \Rm {\mathbb R}

\def \Cm {\mathbb C}

\def \Zm {\mathbb Z}
\def \Sm {\mathbb S}

\newcommand{\dsum}{\displaystyle\sum}
\newcommand{\dint}{\displaystyle\int}
\newcommand{\sspt}{\textsuperscript}

\newcommand{\pdr}[2]{\dfrac{\partial{#1}}{\partial{#2}}}

\newcommand{\bk}{\mathbf k}

\newcommand{\mA}{\mathcal A}

\newcommand{\mC}{\mathcal C}

\newcommand{\mH}{\mathcal H}

\newcommand{\mW}{\mathcal W}

\newcommand{\fC}{{\mathfrak C}}

\newcommand{\fD}{{\mathfrak D}}

\newcommand{\rR}{{\rm R}}

\newcommand{\cout}[1]{}

\newcommand{\sgn}[1]{\,{\rm sign}(#1)}

\newcommand{\op}{\omega_p}
\newcommand{\om}{\Omega}
\newcommand{\ow}{{\rm Op}^w}

\newcommand{\kop}{\pmb{k}} 
\newcommand{\kp}{{\mathbf{k}}} 
\newcommand{\Pha}{{\rm I}}
\newcommand{\Phb}{{\rm II}}
\newcommand{\Phc}{{\rm III}}
\newcommand{\Phd}{{\rm IV}}
\newcommand{\BDI}{{\rm BDI}}

\usepackage{authblk}

\title{Topological edge states of continuous Hamiltonians}

\author[1]{Matthew Frazier}
\affil[1]{Committee on Computational and Applied Mathematics, University of Chicago, Chicago, IL 60637}
\author[1, 2]{ Guillaume Bal \thanks{Corresponding author: \tt guillaumebal@uchicago.edu}}
\affil[2]{Departments of Mathematics and Statistics, University of Chicago, Chicago, IL 60637}

\begin{document}
\maketitle

\begin{abstract}
   This paper concerns the topological classification of continuous Hamiltonians that find applications in biased cold plasmas and photonics. Besides a magnetic bias, the Hamiltonians are parametrized by a plasma frequency and a fixed vertical wavenumber. Eight distinct phases of matter are identified as these parameters vary. When insulating gaps are shared by two such phases, asymmetric edge modes propagate along interfaces separating the two phases. Here we apply the notion of a bulk difference invariant (BDI) to this Hamiltonian, and show by numerical diagonalizations of interface Hamiltonians that after an appropriate regularization our BDI correctly predicts edge transport as described by a bulk edge correspondence. We also derive theoretical tools to compute the BDI and show the limitations of the bulk edge correspondence (BEC) when the phase transition is too singular.
\end{abstract}


\section{Introduction}

Topological systems display phases of transport characterized by topological invariants and, as a consequence, are robust to large classes of perturbations and defects modeled as continuous deformations. A practically and theoretically important example is the topologically protected asymmetric transport occurring along interfaces separating two two-dimensional topological insulators (TI) with different topological characteristics. Such systems find wide applications in electronic structures \cite{hasan2010colloquium,bernevig2013topological}, photonics \cite{PhysRevLett.100.013904,lu2014topological,silveirinha2015chern}, geophysical fluid flows \cite{delplace2017topological}, and cold plasma models \cite{fu2021topological,parker2020topological}. 

The macroscopic description of two-dimensional TI's naturally leads to partial differential equations  on the Euclidean plane. The topological classification of such bulk models poses a number theoretical challenges since the Euclidean plane is not compact  and the standard notion of Chern numbers used to classify tight-binding and periodic materials \cite{bernevig2013topological} or materials displaying a quantum Hall effect with well-defined Landau levels \cite{PhysRevLett.65.2185} does not apply directly. A number of methods have been developed to regularize these systems by means of a one-point compactification mapping the Euclidean plane $\Rm^2$ to the Riemann sphere $\Sm^2\cong \Rm^2 \cup \{\infty\}$ \cite{silveirinha2015chern,shen2012topological,bal2019continuous}. $\Zm-$valued Chern numbers for Hermitian Hamiltonians in Class A are the only ones considered here \cite{bernevig2013topological,prodan2016bulk}. As in \cite{fu2021topological,parker2020topological}, we restrict ourselves to one-dimensional edge models of two-dimensional systems encoded in the bulk-edge correspondence \eqref{eq:BEC} below. For relevant theories in three space dimensions or addressing higher-order topological models, we refer the reader to, e.g., \cite{bernevig2013topological, lin2021real, bal2023topological, benalcazar2017quantized, benalcazar2022chiral}. Their relevance in the analysis of cold plasma models will be analyzed in future studies.

In the classification of the transport asymmetry along interfaces separating different TIs, the notion of bulk phases or bulk invariant is not entirely necessary. One may consider instead the notion of a {\em bulk difference invariant} (BDI), which is defined in greater generality. BDI were introduced in \cite{bal2022topological} and further analyzed in \cite{quinn2024approximations,rossi2024topology,bal2024continuous}. They confirm the intuition that phase differences are more generally defined than differences of absolute phases. They were applied in the context of electronic structures to classify interface Hamiltonians modeling bilayer graphene problems \cite{bal2023mathematical} and models of Floquet topological insulators \cite{bal2022multiscale}.  The main technical step in the construction of the BDI is to use a radial compactification of the two Euclidean planes modeling wavenumber of the two phases, mapping them to hemispheres of a unit sphere and gluing them continuously along the equator. 

The primary usefulness of BDI is their correct prediction of asymmetric edge transport by means of a {\em bulk-edge correspondence} (BEC), a central physical principle in our understanding of topological phases of matter \cite{bernevig2013topological,prodan2016bulk,hatsugai1993chern,essin2011bulk,silveirinha2016bulk}. The BEC states that the excess of number of modes propagating in one direction along the interface with respect to the number propagating in the opposite direction is given by the topological phase difference, i.e., the BDI. 

To classify asymmetric interface transport, the notion of excess of number of modes is somewhat imprecise unless a complete diagonalization of the interface Hamiltonian is available. A more general notion is the interface current observable $\sigma_I$ described in \eqref{eq:sigmaI} below. The interface current, which may be computed by {\em spectral flow} of an interface Hamiltonian with known diagonalization \cite{bal2024continuous}, was introduced in \cite{schulz2000simultaneous} and used in many mathematical derivations of the bulk edge correspondence \cite{prodan2016bulk,elbau2002equality,drouot2021microlocal}. In particular, the BEC in  \cite{quinn2024approximations, bal2023topological} takes the form $2\pi\sigma_I=\,$BDI for {\em elliptic} partial differential operators. Ellipticity is a property of bulk and interface Hamiltonians essentially implying that the energy or frequency is large for wavepackets with large wavenumber, and this uniformly in the position of the wavepacket in the Euclidean plane; see \cite{quinn2024approximations,bal2024continuous,bal2023topological} for a more precise definition.

Our analysis is carried out for a specific $9\times9$ Hermitian Hamiltonian modeling a linearization of a coupling between a macroscopic electron current with an electromagnetic field in the presence of a magnetic bias responsible for the non-trivial topological properties.  
This system contains multiple flat bands at large wavenumbers, and therefore is not elliptic. Be that as it may, we wish to study the BEC in the context of BDI's and see whether the BEC can still apply in such non-elliptic cases.
In particular, whereas appropriate high-wavenumber regularizations of photonic models led to integer-valued integrals of Berry curvatures, the BEC was not always satisfied \cite{silveirinha2016bulk,hanson2016notes,pakniyat2022chern,qin2023topological,serra2025influence}, and indeed we find that these integer-valued integrals of Berry curvature do not always define BDI's of the system.  The main objective of this paper is to analyze how the construction of BDI helps to restore the BEC. In particular, we introduce regularizations of photonics and cold plasma models in which the aforementioned gluing of the two phases may be obtained continuously so as to define a {\em bona fide} Chern number for families of projectors parametrized by the unit sphere.

Our analysis proceeds as follows: We perform a systematic theoretical analysis of all possible topological phases of the system and all possible frequency gaps shared by two different phases. We then construct BDI for two such phases and show with numerical diagonalization of the corresponding interface Hamiltonian gluing these two phases that the BDI correctly predicts a BEC for most transitions. For phase transitions that are a priori feasible since they share a common insulating gap but involve singular variations in the Hamiltonian, we demonstrate with simple $2\times2$ systems why the BEC should not be expected to hold. By analogy to electronic structures, the type of singularities we address correspond to Hamiltonians with locally vanishing Fermi velocities. We provide a detailed analysis of why the notion of BEC breaks down in such cases even when the Hamiltonians involve smooth coefficients and no edges.

The rest of the paper is structured as follows. Section \ref{sec:invariants} recalls the appropriate notions of invariants for two dimensional bulk and interface Hamiltonians in class A, in particular interface current observables, integrals of Berry curvature, Chern numbers obtained by one-point compactification and by radial compactification, and a simple method to compute such integrals theoretically for rotationally symmetric Hamiltonians. Section \ref{sec:model} describes our cold plasma model and its topological classification, essentially building on the past works \cite{fu2021topological,parker2020topological, silveirinha2016bulk,hanson2016notes,qin2023topological}. 
Our main results are presented in Section \ref{sec:main}.
The central novelty of our analysis is the introduction of high-wavenumber regularizations of the Hamiltonian that differ from those in the aforementioned references and allow us to define {\em bona fide} integer-valued BDI. Extensive numerical simulations of interface Hamiltonians modeling the spatial transition between different topological phases prove that our BDI is consistent with the BEC in all cases but one. Finally, Section \ref{sec:disc} concludes with novel material presenting simple Dirac-type Hamiltonians for which the BEC should not be expected to hold. We show that in the one case where BDI's fail to validate the BEC in the cold plasma model, the interface Hamiltonian can be reduced to one such highly singular Dirac-type Hamiltonian.  Additional details on derivations and numerical methods may be found in Appendices.

\section{Interface and Bulk invariants}
\label{sec:invariants}

This section summarizes the construction and computation of bulk and interface invariants that we will use in the next section. We refer to \cite{bernevig2013topological,silveirinha2015chern,prodan2016bulk,bal2024continuous} for additional detail.

We consider continuous Hermitian Hamiltonians $H_I$ acting on vector-valued functions in two space dimensions $(x,y)\in\Rm^2$ and modeling a smooth transition  near the {\em equator} $\{y=0\}$ between two topological insulators denoted by constant-coefficient Hamiltonians $H^h$ for $h\in\{N,S\}$. We assume that the coefficients of $H_I$ are given by those of $H^N$ when $y\geq1$ and by those of $H^S$ when $y\leq-1$. The main example of Hamiltonian considered here is the light-matter interaction system in \eqref{eq:HI} below.

We use the following notation.
We denote by $\kp=(k_x,k_y)$ the conjugate (Fourier) variables to $(x,y)$. We denote by $\Pi_j^h$ or $P_j^h$ families of projectors related to the band structure of $H^h$. The integrals of Berry curvatures associated to a projector $\Pi$ are denoted by $\mC[\Pi]$. They do not need to be integer-valued. Bona fide integer-valued Chern numbers are systematically denoted by $\fC[\cdot]$. We use $\bar \alpha$ to denote complex conjugation of $\alpha$ and $A^*$ to denote Hermitian conjugate of an operator $A$.
 
\subsection{Interface invariant}

A robust and general way to quantify the interface asymmetry is to consider the following interface current observable \cite{prodan2016bulk,bal2024continuous,elbau2002equality}.

Let $P=P(x)$ be a function that depends only the spatial coordinate $x$ with  $P(x)=0$ for $x<x_0-\delta$ and $P(x)=1$ for $x>x_0+\delta$ for some $x_0\in\Rm$ and $\delta>0$. The function $P(x)$ should be interpreted as the observable quantifying the field density in the (right) half-space $x\geq x_0$. Then $i[H,P]$ with $[A,B]=AB-BA$ the standard commutator, may naturally be interpreted as a current operator modeling transfer of field density  per unit time across the (thick) vertical line where $P$ transitions from $0$ to $1$.

The bulk Hamiltonians $H^N$ and $H^S$ are both assumed to be insulators for a frequency or energy interval $[E_0,E_1]$. Any excitation generated in such an interval will therefore be confined to the vicinity of the interface $y\approx0$ separating the insulators. We define $0\leq\varphi\in C^\infty(\Rm)$ as a function such that $\varphi(E)=0$ for $E\leq E_0$ and $\varphi(E)=1$ for $E\geq E_1$. Thus $\varphi'(H_I)$ defines a density of states of modes that cannot propagate into the N and S bulks. 

The expectation value of the current observable $i[H_I,P]$ for a density of states $\varphi'(H_I)$ is then defined as:
\begin{equation}\label{eq:sigmaI}
  \sigma_I [H_I] = {\rm Tr} \ i[H_I,P] \varphi'(H_I)
\end{equation}
assuming that $i[H_I,P] \varphi'(H_I)$ is a trace-class operator. Here, ${\rm Tr}$ is the standard trace on the Hilbert space $\mH$ where $H_I$ is defined. We will refer to $\sigma_I$ as an interface (current) observable. It is the physical object describing asymmetric transport along an interface, which here is modeled as the $x-$axis.

This observable has been analyzed and related to topological invariants in many settings; see e.g. references in \cite{bal2024continuous}. The computation of the interface current observable is difficult in practice unless the operator $H_I$ may be fully diagonalized; see \cite[\S3]{bal2024continuous} for a computation using spectral flows, which is how $2\pi\sigma_I$ will be computed numerically in section \ref{sec:main}. Its computation is significantly simplified if it may be related to the properties fo the bulk Hamiltonians $H^N$ and $H^S$. A general principle, the bulk-edge correspondence (BEC) allows one to do so. We first recall the construction of bulk invariants.

\subsection{Bulk and Bulk difference invariants (BDI)}

\begin{figure}[t!]
    \centering
    \includegraphics[width = \textwidth]{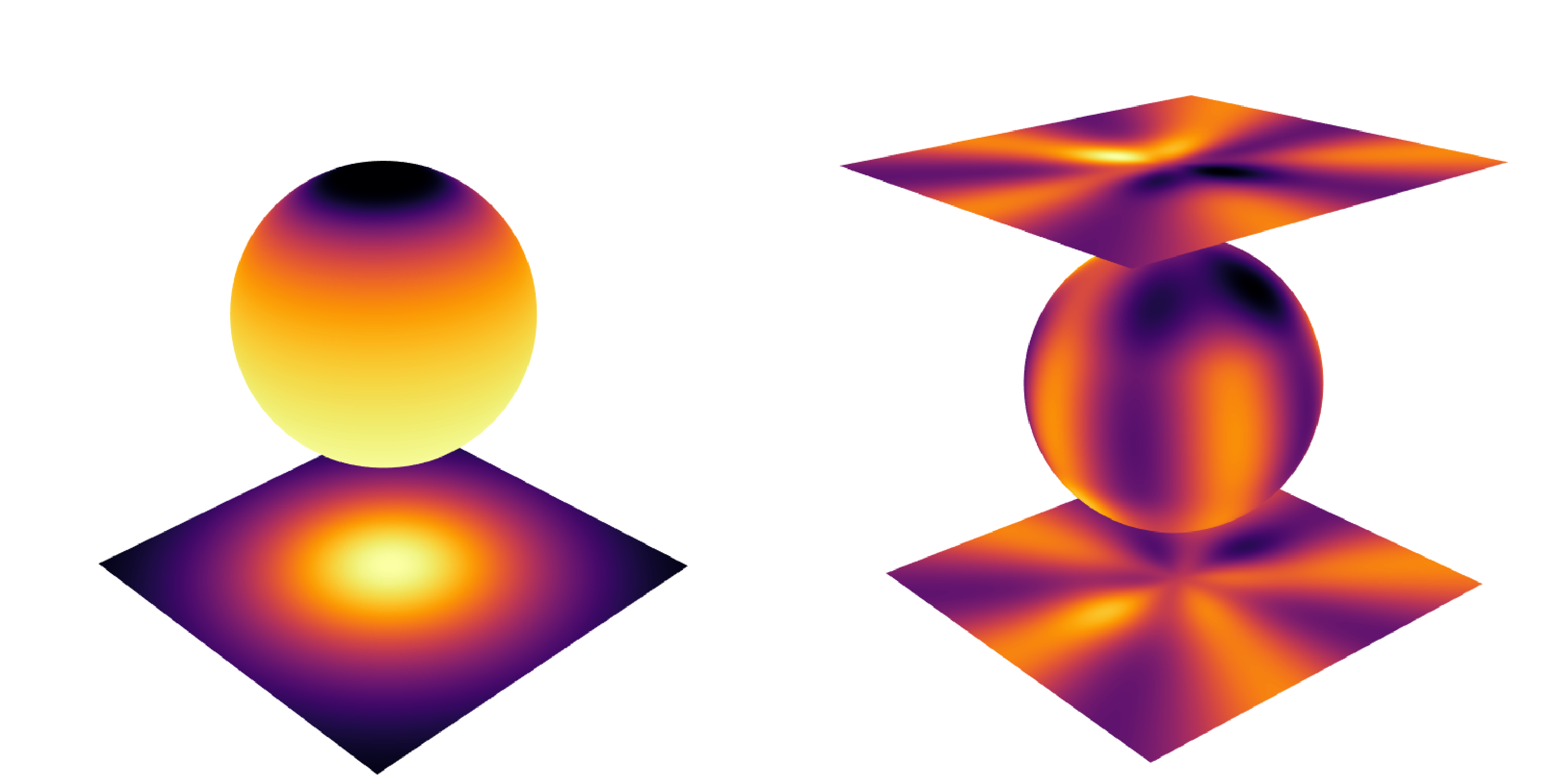}
    \caption{Visual comparison between one-point (left) and radial (right) compactification. On the left the function on $\Rm^2$ mapped onto $\Sm^2$ is continuous due to the limit at infinity being uniform in all directions. Meanwhile on the right two functions on $\Rm^2$ are mapped to the upper- and lower-half of $\Sm^2$ respectively and are glued together along the equator. As long as condition \eqref{eq:gluing} holds the combination is continuous on $\Sm^2$.}
    \label{fig:compact}
\end{figure}

We consider two families of self-adjoint Hamiltonians in Fourier variables $H^h(\kp)$ for $h\in\{N,S\}$ and $\kp\in\Rm^2$ with values in $\Cm^n\times\Cm^n$ and assume the following spectral decomposition
\[  H^h(\kp) = \dsum_{j=1}^n \lambda_j^h(\kp) \Pi^h_j(\kp)\]
where $\Pi^h_j(\kp)=\psi^h_j(\kp)\otimes\psi^h_j(\kp)$ are rank-one projectors and $\lambda^h_j(\bk)$ are the corresponding eigenvalues. Associated to each (arbitrary-rank) projector family $\Rm^2\ni \kp\mapsto\Pi(\kp)$ is the following integral of the associated (Berry) curvature
\begin{equation}\label{eq:curv}
    \mC[\Pi]=\frac i{2\pi} \int_{\Rm^2} {\rm tr} \Pi d\Pi \wedge d\Pi,\qquad 
d\Pi := \pdr{\Pi}{k_x}dk_x + \pdr{\Pi}{k_y}dk_y.
\end{equation}
Here ${\rm tr}$ stands for standard matrix trace.
Because $\Rm^2$ is not a compact manifold, the above integral is not necessarily a Chern number nor is it necessarily integral-valued.

\paragraph{One-point compactification  and absolute phase invariants.} 
When the projector family $\Pi(\kp)$ has a unique limit at the point at infinity $\kp=\infty$, i.e., when $\Pi(r\theta)\to \Pi_\infty$ as $r\to\infty$ with $\Pi_\infty$ independent of $\theta\in\Sm^1$, we may then project the plane $\Rm^2$ stereographically onto the Riemann sphere $\Sm^2$, its one-point compactification. The resulting family of operators on the Riemann sphere is then continuous, in which case we define
\begin{equation}\label{eq:Chernop}
    \fC[\Pi] := \mC[\Pi] \in \Zm
\end{equation}
as a {\em bona fide} Chern number with base manifold the Riemann sphere $\Sm^2$ \cite{silveirinha2015chern,bal2019continuous}.

\paragraph{Radial compactification and bulk-difference invariant.}

In many applications, the family of projectors is not continuous at the point at infinity $\kp=\infty$. A radial compactification may be preferred instead.

Assume a spectral gap between levels $\ell$ and $\ell+1$ for both $h=N$ and $h=S$, i.e., a frequency interval $I_\ell$ such that $\lambda^h_j(\kp)< I_\ell$ for $h\in{N,S}$ and $j\leq \ell$ while $\lambda^h_j(\kp)> I_\ell$ for $h\in{N,S}$ and $j \geq \ell+1$. Associated to the spectral gap are the projectors:
\[
   P_\ell^h = \sum_{j\leq \ell} \Pi^h_j,\qquad \mW_\ell^h := \mC[P_\ell^h] = -\mC[I-P_\ell^h],\qquad I-P_\ell^h = \sum_{j\geq \ell+1} \Pi^h_j.
\]
The total curvature associated to a band may be computed as a sum either over bands below the gap or over bands above the gap. We will use the latter option in section \ref{sec:model}.
We then define the bulk-difference invariant for the gap labeled by $\ell$:
\begin{equation}\label{eq:ChernBDI}
    \fC_\ell := \fC[P_\ell^S,P_\ell^N] := \mW_\ell^S - \mW_\ell^N  = \frac  i{2\pi} \int_{\Rm^2} {\rm tr} P_\ell^S dP_\ell^S\wedge dP_\ell^S
 - \frac  i{2\pi} \int_{\Rm^2} {\rm tr} P_\ell^N dP_\ell^N \wedge dP_\ell^N.
\end{equation}

Provided that we have the following gluing condition 
\begin{equation}\label{eq:gluing}
    \lim_{r\to\infty} P_\ell^N(r\theta) = \lim_{r\to\infty} P_\ell^S(r\theta) \quad \mbox{ for all } \theta\in \Sm^1,
\end{equation}
then $\fC_\ell$ is also a {\em bona fide} Chern number for a family of projectors defined on the sphere $\Sm^2$. Indeed, we stereographically project $P^N_\ell$ onto the upper hemisphere of $\Sm^2$ and $P^S_\ell$ onto the lower hemisphere of $\Sm^2$ while the above gluing condition ensures that the family of projectors on $\Sm^2$ is continuous across the equator. This guarantees that $\fC_\ell\in\Zm$, which we call a {\em Bulk Difference Invariant} (BDI); see Figure \ref{fig:compact}.

In practice, we often verify the following continuity condition for each band $j\leq \ell$, 
\[
  \lim_{r\to\infty} \Pi_j^N(r\theta) = \lim_{r\to\infty} \Pi_j^S(r\theta) \quad \mbox{ for all } \theta\in \Sm^1.
\]
This implies that 
\[
  \fC[\Pi_j^S,\Pi_j^N] = \mC[\Pi_j^S]- \mC[\Pi_j^N] \in \Zm,\qquad 
   \fC[P_\ell^S,P_\ell^N]  = \dsum_{j\leq \ell} \fC[\Pi_j^S,\Pi_j^N],
\]
by the (non-trivial) additivity property of Chern numbers. Therefore, in practice, to define a BDI we need only verify that \eqref{eq:gluing} holds and then sum the integrals of Berry curvature. In this way our computation of BDI's is similar to the calculation of Chern numbers in \cite{silveirinha2015chern, fu2021topological, silveirinha2016bulk, hassani2016effects, hanson2016notes}, however most of our efforts will be in identifying conditions under which \eqref{eq:gluing} holds, which was not addressed previously.

That the radial compactification is more generally defined than the one-point compactification reflects the fact that topological phase differences are more generally defined than absolute topological phases. In the application to continuous models in photonics and cold plasmas, unperturbed models display behavior of projectors $\Pi(\kp)$ as $|\kp|\to\infty$ that allow for neither one-point nor radial compactification. We will consider several {\em high-wavenumber regularizations} \cite{silveirinha2015chern, parker2020topological,silveirinha2016bulk,hanson2016notes,pakniyat2022chern,qin2023topological} and show that those regularizations for which BDI can be constructed correctly predict the presence and number of topologically protected edge states.

One interesting note is that the integral of Berry curvature $\mathcal{C}[\Pi]$ can be an integer without satisfying either one-point or radial compactification  \cite{silveirinha2015chern}. The central contribution of \cite{silveirinha2015chern} was proposing a more flexible framework to apply Chern invariants to continuum systems by simply requiring that $\mathcal{C}[\Pi_j] \in \mathbb{Z}$ for all bands of the system using high-wavenumber regularizations. However, one of our central findings, detailed in Section 4, is that some natural high-wavenumber regularizations which produce integer integrals of Berry curvature do so in spite of the fact that neither one-point nor radial compactification apply, and therefore do not define BDI's for the system. Our numerical results suggest that correct predictions of topological edge states occur when well-defined BDI's are used, regardless of the method of regularization.

\subsection{Computation of Chern numbers in isotropic case.}

The computation of the above invariants involves integrals of the form $\mC[\Pi]$, whether they are integral-valued or not. Computing such integrals analytically is not always straightforward but drastically simplifies in the presence of rotational symmetry. 
Assume $\Pi(\kp)=\psi(\kp)\psi^*(\kp)$ a rank-one projector. We then verify that
\[
 {\rm tr} \Pi d\Pi \wedge d\Pi = d A,\qquad A(\kp) = (\psi(\kp),d\psi(\kp))
\]
where $A$ is a one-form-$i\Rm$-valued (Berry) connection assuming $d\psi(\kp)$ continuously defined. As an application of the Stokes theorem, we thus observe \cite{bernevig2013topological,silveirinha2015chern,hanson2016notes} that
\begin{equation}\label{eq:Stokes}
     \mC[\Pi] = \frac{i}{2\pi}\Big(\oint_{|\kp| \rightarrow \infty} A(\kp)  - \oint_{|\kp|\rightarrow  0}  A(\kp)\Big).
\end{equation}
When $A(\kp)$ is continuously defined at $\kp=0$, then the above integral over circles with vanishingly small radii converges to $0$. However, writing $\kp=|\kp|e^{i\theta}$ for $\theta\in[0,2\pi)$, we observe that the above formula remains valid if $\psi(\kp)$ is (globally) gauge-transformed to, e.g., $e^{im\theta}\psi(\kp)$. This flexibility proves convenient in practice.

Assume now that the Hamiltonian family is isotropic. Let $\kp=ke^{i\theta}$ for $e^{i\theta}\in\Sm^1$ the unit circle. We assume that 
\[
H(k\theta) = H(\theta) = U(\theta) H(0) U^*(\theta)
\]
for $U(\theta)$ a family of $\Cm^n-$ unitary transformations. This implies that the branches of spectrum $\lambda_j(\kp)=\lambda_j(k)$ are independent of $\theta$ and we may choose the eigenvectors $\psi_j(\theta)=U(\theta)\psi_j(0)$. 

Isotropy, or invariance by rotation, implies that $U^*(\theta)dU(\theta)= U'(0)d\theta$ is independent of $\theta\in[0,2\pi)$. In this setting, we thus obtain that
\begin{equation}\label{eq:simpChern}
    \mC[\Pi] = i\big[ \lim_{k\to\infty} (\psi(k),U'(0)\psi(k)) - \lim_{k\to0} (\psi(k),U'(0)\psi(k)) \big].
\end{equation}
In other words, all we need to compute is $\psi(\kp)$ for $\kp=(k,0)$ with $k=0$ and $k=\infty$. Explicit expressions for the projectors as $k\to\infty$ are also necessary in order to verify gluing conditions \eqref{eq:gluing}. For the cold plasma application, such computations are obtained analytically. 

\subsection{Bulk-Edge Correspondence}
\label{sec:BEC}

The bulk-edge correspondence is a general principle stating that the edge current asymmetry $2\pi\sigma_I$ is related to the bulk invariants by the relation
\begin{equation}\label{eq:BEC}
 2\pi\sigma_I[H_I] = \BDI = \fC[P^S_\ell,P^N_\ell]
\end{equation}
where $\ell$ is a common spectral gap of the bulk Hamiltonians $H^h$ for $h\in\{N,S\}$ and the density $\varphi'$ appearing in \eqref{eq:sigmaI} is supported in that common gap. This relation thus implies that the number of edge modes characterized by $2\pi\sigma_I$ is independent of the details of the transition between $H^S$ and $H^N$.

While natural and ubiquitous in the analysis of topological phases of matter \cite{bernevig2013topological,prodan2016bulk}, the BEC does not always hold for continuous Hamiltonians. A class of continuous operators for which it is guaranteed to apply is that of {\em elliptic} operators \cite{bal2022topological,bal2024continuous,quinn2024approximations}. Elliptic operators are essentially characterized by singular values $|\lambda^h_j(\kp)|\to\infty$ as $|\kp|\to\infty$ for all branches $j$ and both $h\in\{N,S\}$; see above references. 

The operators naturally appearing in shallow water models \cite{delplace2017topological}, photonics \cite{silveirinha2015chern,pakniyat2022chern}, or biased cold plasmas \cite{parker2020nontrivial,qin2023topological} are not elliptic and it is not always immediate how to naturally regularize them so that they become elliptic. For the well-studied shallow water problem, the BEC is shown to fail in the presence of non-smooth coefficients or boundary conditions \cite{graf2021topology,bal2024topological,graf2024boundary}. Yet natural elliptic regularizations, for which the BEC holds, may be introduced \cite{bal2022topological,quinn2024approximations,bal2024continuous}. This is more difficult to implement in the context of photonics and cold plasma models since the obstruction to ellipticity involves several topologically non-trivial bands. In section \ref{sec:main}, we will aim to regularize the Hamiltonians to ensure that we have defined invariants $\BDI = \fC[P^S_\ell,P^N_\ell]$. That \eqref{eq:BEC} still holds in that case will be demonstrated with numerical simulations of $2\pi\sigma_I$ using spectral flow analyses.

To analyze several phase transitions more precisely, we will introduce reduced Hamiltonians as was done in, e.g., \cite{qin2023topological}. These reduced Hamiltonians model the interaction of two bands, and as such take the form of Dirac-type operators. Reduced interface operators $\tilde H_I$ are then either elliptic or singular depending on the phase transition of interest. For elliptic operators, the theory of \cite{quinn2024approximations,bal2024continuous,bal2023topological} validates \eqref{eq:BEC}. For singular operators, we demonstrate in Section \ref{sec:disc} that \eqref{eq:BEC} does not hold.

\paragraph{Reduced Hamiltonian.} Consider a bulk Hamiltonian $H$ with two bands associated to eigenvectors $\psi_j(\kp)$ for $j=1,2$ crossing at $\kp=0$ without loss of generality at a frequency $\lambda_1(0)=\lambda_2(0)$. We may then define the reduced Hamiltonian
\[
 \fD_{ij} = (\psi_i,H\psi_j),\quad 1\leq i,j,\leq 2.
\]
This reduced Hamiltonian takes the form of a two-band Dirac-type operator; see next section and \cite{qin2023topological}. After formally introducing spatial variations in the constitutive coefficients of the operator, we consider the following model
\begin{equation}\label{eq:Diracgal}
    \fD_I =  v_x(y) D_x \sigma_1 + \frac12\{v_y(y),D_y\} \sigma_2 + [m(y)+\eta\Delta] \sigma_3 + V(y)
\end{equation}
where $\sigma_{1,2,3}$ are the standard Pauli matrices, $D_{x,y}=-i\partial_{x,y}$, $\{A,B\}=AB+BA$ (ensuring that the above operator is self-adjoint), and where the velocity coefficients $v_{x,y}$, the mass term $m(y)$, and the potential term $V(y)$ are smooth functions allowed to depend on $y$. Here, $\eta$ is a small regularization term and $\Delta$ is the usual Laplace operator.

When $v_x$ and $v_y$ are constant and $V(y)=0$, then the BEC \eqref{eq:BEC} holds for the above model \cite{quinn2024approximations, bal2023topological}, of course assuming $\fD^N$ and $\fD^S$ are bulk Hamiltonians with a common gap and $\varphi'$ in \eqref{eq:sigmaI} is supported in that gap. The operator is indeed elliptic both before ($\eta=0$) and after ($\eta\not=0$) regularization. When $\eta=0$ and $v_x(y)$ or $v_y(y)$ change signs, however, the operator $\fD_I$ is no longer elliptic. We will call such Hamiltonians singular. These singularities generate violations to the bulk-edge correspondence of a more serious type than for the aforementioned shallow water model \cite{quinn2024approximations,graf2021topology,bal2024topological,graf2024boundary}, which are discussed as they pertain to the cold plasma model in Section \ref{sec:disc}.

\section{Topological phases in cold plasma model}
\label{sec:model}

\subsection{Cold plasma Hamiltonian model}
\label{sec:cph}

We now focus on the analysis of a continuous Hamiltonian that finds applications in the modeling of biased cold plasmas and photonic materials \cite{silveirinha2015chern,qin2023topological,parker2021topological}.
We start with the following linearized light-matter interaction system
\begin{equation}
    \begin{array}{rcl}
        m_e \pdr vt & = & q_e(E + v\times B_i)  \\[2mm]
        c^2 \nabla \times B & = & 
	\dfrac{ n_e q_e v}{\epsilon_0} + \pdr Et
    \\[3mm]
    \nabla \times E &=& -\pdr Bt,
    \end{array}
\end{equation}
where $v(\mathbf{r})$ denotes electron current and $(E(\mathbf{r}),B(\mathbf{r}))$ the electromagnetic field at $\Rm^3\ni \mathbf{r}=(x,y,z)$. Also, $(m_e,q_e,\epsilon_0)$ are fundamental constants while $B_i=B_0(x,y)\hat{e}_z$ and $n_e=n_e(x,y)$ are given bias magnetic field and plasma density, respectively. Assuming that the Lorentz force depends only on the incident magnetic field amounts to assuming that to 0-th order the electron velocity is 0, $v^{(0)} = 0$, which is the main assumption of the cold plasma model. See \cite{stix1992waves} for more discussion and a derivation as a limit of kinetic theory of hot plasma. 
We thus assume the system invariant by translations along the $z$ variables so that the dual variable $k_z$ is a good quantum number that becomes a parameter in the system. Note that in real applications the $z-$direction is always finite, and the effects of broken translation-invariance in this case are treated extensively by numerical studies in \cite{hassani2016effects}.

We introduce the rescaled quantities and change of variables
\[\Omega(x,y) = \frac{q_e B_0(x,y)}{m_e},\ \omega_p(x,y) =  \sqrt{\frac{n_e(x,y)q_e^2 }{m_e\epsilon_0}},\ \ cB\to B,\ v(t,x,y)\sqrt{\frac{m_e n_e(x,y)}{\epsilon_0}} \to v(t,x,y)\]
where $\Omega(x,y)$ is the cyclotron frequency and $\omega_p(x,y)$ the plasma frequency, to obtain the Hamiltonian system of $9\times9$ equations for $\psi=(v,E,H)^T$:
\begin{equation}\label{eq:HI}
    i\partial_t \psi = H_I\psi,\qquad H_I:=\begin{pmatrix} i\Omega(x,y) \hat{e}_z\times & -i\omega_p(x,y) & 0   \\
     i\omega_p(x,y) & 0 & i\nabla\times \\
     0 & -i\nabla \times & 0\end{pmatrix}.
\end{equation}
We denote by $H_I$ this Hamiltonian to include a transition between different insulating bulks and will make use of the generic cross-product operator in matrix form: 
\[
u \times := \begin{pmatrix} 0 & -u_z & u_y  \\
     u_z & 0 & -u_x \\
     -u_y & u_x & 0\end{pmatrix},\qquad u:=(u_x,u_y,u_z)^t.
\]
In this paper, we assume all parameters of the system independent of $x$ and continuous in $y$. The above Hamiltonian defines the constant coefficient Hamiltonians $H^N$ when $y\gg1$ and $H^S$ when $y\ll\-1$. In the dual Fourier variables $\kop=(\kp,k_z)$ to $\mathbf{r}$, these operators take the form for each fixed $k_z\in\Rm$ of the family
\begin{equation}\label{eq:HFourier}
  \Rm^2\ni \kp\mapsto  H(\kp) = \begin{pmatrix} i\Omega(k)\hat{e}_z\times & -i\omega_p(k) & 0   \\
     i\omega_p(k) & 0 & -\kop\times \\
     0 & \kop\times & 0\end{pmatrix}.
\end{equation}
While the coefficients $(\Omega,\omega_p)$ are constant in the model, we will assume the presence of regularization mechanisms to suppress unwanted phenomena as $\kp\to\infty$. This is anticipated in the above expressions where $\Omega(k)$ and $\omega_p(k)$ are allowed to depend on $k=|\kp|$.

Our objective in this section is to associate interface invariants to $H_I$ and bulk difference invariants to $H(\kp)=H(\kp;\Omega,\omega_p,k_z)$ as the parameters $(k_z,\Omega,\omega_p)$ vary.

\subsection{Symmetries of the problem}
\label{sec:symmetries}
We first analyze the spectral properties of the bulk Hamiltonians $H(\kp)=H(\kp;\Omega,\omega_p,k_z)$. 

\paragraph{Invariance by rotation.} We decompose $\kp=ke^{i\theta}$ for $\theta\in[0,2\pi)$ and introduce $R(\theta)$ be the family of orthogonal operators
\[ R(\theta) = \begin{pmatrix} \cos\theta & -\sin\theta& 0 \\ \sin\theta & \cos\theta& 0 \\ 0 & 0 & 1\end{pmatrix}\quad \mbox{ such that } \quad \kop = R(\theta) (k,0,k_z)^T.
\]
Using the notation $H(\theta)=H(ke^{i\theta})$ and
defining $\rR(\theta) = {\rm Diag} (R(\theta),R(\theta),R(\theta))$, we observe that
\begin{equation}\label{eq:invrot}
    \rR^*(\theta) H(\theta) \rR(\theta) = H(0).
\end{equation}
This is a direct consequence of the fact that $\kop\times$ and $e_z\times$ satisfy the above invariance. This shows that the spectrum of the family $H(\kp)$ is invariant by rotation parametrized by $\theta\in[0,2\pi)$. Moreover, the transformation of eigenvectors is given by $\psi(\theta)=\rR^*(\theta)\psi(0)$ and \eqref{eq:simpChern} applies with $U'(0)=\rR'(0)$.

\paragraph{Discrete symmetries.}
Define the parity operator
\[\Gamma_P(\theta) = {\rm Diag} (S(\theta),-S(\theta),-S(\theta)),\ 
 S(\theta) = \begin{pmatrix} \cos(2\theta)\sigma_3 + \sin(2\theta)\sigma_1 & 0 \\ 0 & 1\end{pmatrix} = R(\theta) S(0) R^*(\theta).\]
We observe that $\Gamma_P(\theta)=\Gamma^*_P(\theta)$ and $\Gamma_P^2(\theta)=I_9$. Moreover, we have the parity invariance
\begin{equation}\label{eqparityinv}
    \Gamma_P(\theta) H(\theta) \Gamma^*_P(\theta) = -H(\theta)
\end{equation}
This shows that $H(\theta)$ is unitarily equivalent to $-H(\theta)$. This parity relation implies that the spectrum of $H$ is symmetric about $0$. Moreover, for eigenvalues different from $0$, their corresponding eigenvectors $\psi(\theta)$ and $\Gamma^*_P(\theta)\psi(\theta)$ are orthogonal. 

This result comes from an explicit calculation where $k_y$ is first sent to $0$ by rotation and then the observation that $S(0) k\times S(0) = -k\times$ when $k_y=0$. We observe for the same reason that $S(\theta) \hat e_z\times S(\theta)=-\hat e_z$.

Defining the operators $\Gamma_\Omega= {\rm Diag} (I_3,-I_3,I_3)$ and $\Gamma_k= {\rm Diag} (I_3,I_3,-I_3)$, we observe that 
\begin{equation}\label{eq:symH}
    \Gamma_\Omega H(\theta;\Omega)\Gamma_\Omega = - H(\theta;-\Omega),\qquad
\Gamma_k H(\theta;k)\Gamma_k =  H(\theta;-k).
\end{equation}
Since $\omega_p>0$ physically, we do not consider the case $\omega_p<0$.

Combining the two preceding results shows that the spectrum is invariant under $\Omega\to-\Omega$:
\begin{equation}\label{eq:invOmega}
    \tilde\Gamma_P(\theta) H(\theta,\Omega) \tilde\Gamma_P(\theta) = H(\theta,-\Omega), \qquad \tilde\Gamma_P(\theta)= \tilde\Gamma_P^*(\theta) :=\Gamma_P(\theta)\Gamma_\Omega.
\end{equation}
For $H(\theta,\Omega)\psi=\omega\psi$, we thus obtain that $H(\theta,-\Omega)[\Gamma^*_P(\theta)\Gamma_\Omega\psi]=\omega[\Gamma^*_P(\theta)\Gamma_\Omega\psi]$. Note that $\tilde\Gamma_P(\theta)={\rm Diag} (S(\theta),S(\theta),-S(\theta))$.

Similarly, defining $\tilde \Gamma_k=\Gamma(k) \rR(\pi)$, we observe that 
\begin{equation}\label{eq:invkz}
    \tilde \Gamma_k H(\kp;k_z) \tilde \Gamma_k= H(\kp;-k_z)
\end{equation}
so that for instance $H(k_z)\psi=\omega\psi$ implies that $H(-k_z) [\tilde \Gamma_k\psi] = \omega[\tilde \Gamma_k\psi]$.

Note that $\bar H(k) = - H(-k)$ as one readily verifies so that $\Gamma_k \bar H(k)\Gamma_k=-H(k)$. Note also that $\bar\psi(k)=\psi(-k)$ since $H_I$ has real-valued coefficients in the physical variables. This imposes that $H(k) (\Gamma_k\psi(-k)) = -\omega(k) (\Gamma_k\psi(-k))$
also implying that the spectrum of $H$ is symmetric about $0$ as shown in, e.g., \cite{qin2023topological}. The above derivation defines a precise unitary equivalence between $H(k)$ and $-H(k)$ via $\Gamma_P(\theta)$.

\subsection{Topological phases}
\label{sec:phases}
We saw in the preceding section that all we needed to compute integrals as in \eqref{eq:curv} were the eigenvectors $\psi(k,0)$ for $k\in\Rm_+$. In order to be able to apply Stokes formula leading to \eqref{eq:Stokes} we make the assumption that the branches $k\to\lambda_j(k)$ do not cross. We do not have a theoretical justification for this absence of crossing for $0<k<\infty$ beyond extensive numerical simulations. In fact, branches do cross at $k=0$, $k=\infty$, $k_z = 0$, and $\om = 0$. This is the reason for the existence of several phases of matter for the cold plasma model, and why first we consider only strictly positive $(\om, \op, k_z)$ and treat the $k_z = 0$ and $\om = 0$ cases separately.

By parity invariance, $\omega=0$ is always an eigenvalue of $H(\kp)$, which we verify is associated to $\psi_0=(0,0,\kop/|\kop|)$.

We first consider the rest of the diagonalization of $H(k=0)$, assuming all coefficients $(\Omega,\omega_p,k_z)$ strictly positive. By parity, there are four positive eigenvalues and four negative eigenvalues. The four positive eigenvalues are denoted by $(\omega_p,\omega_{L-},\omega_R,\omega_{L+})$.

Indeed we verify that $\omega=\omega_p$ is always an eigenvalue associated to the eigenvector $\Psi_p=2^{-\frac12}(\hat e_z,i\hat e_z,0)^T$. The other eigenvalues are the positive solutions of two cubic equations given explicitly by
\begin{equation}\label{cubicpm}
    k_z^2 = \omega_R^2 - \frac{\omega_p^2\omega_R}{\omega_R+\Omega},\qquad k_z^2 = \omega_L^2 - \frac{\omega_p^2\omega_L}{\omega_L-\Omega}.
\end{equation}
Associated to these eigenvalues are the eigenvectors, up to normalizing constants $c$:
\[
\begin{aligned}
    \Psi_{L-} = c\begin{pmatrix}
        i\frac{\omega_p}{\Omega - \omega_{L^-}} \hat{e}_+\\ \hat{e}_+\\-i\frac{k_z}{\omega_{L^-}} \hat{e}_+
    \end{pmatrix},\ 
    \Psi_R = c\begin{pmatrix}
        - i\frac{\omega_p}{\Omega+\omega_R} \hat{e}_- \\ \hat{e}_- \\i\frac{k_z}{\omega_R} \hat{e}_- 
    \end{pmatrix},\
    \Psi_{L+} = c\begin{pmatrix}
        i \frac{\omega_p}{\Omega - \omega_{L^+}} \hat{e}_+ \\
         \hat{e}_+ \\
         -i\frac{k_z}{\omega_{L^+}} \hat{e}_+
    \end{pmatrix},\ \hat{e}_\pm = \frac{1}{\sqrt2}\begin{pmatrix}
        1 \\ \pm i \\ 0
    \end{pmatrix}.
\end{aligned}
\]
We always have $0<\omega_{L-}<\omega_R<\omega_{L_+}$ (see Appendix F for details) but the relation among the other eigenvalues is not fixed and in fact defines different topological phases.

\begin{figure}[ht]
			\centering
			\includegraphics[width=\textwidth]{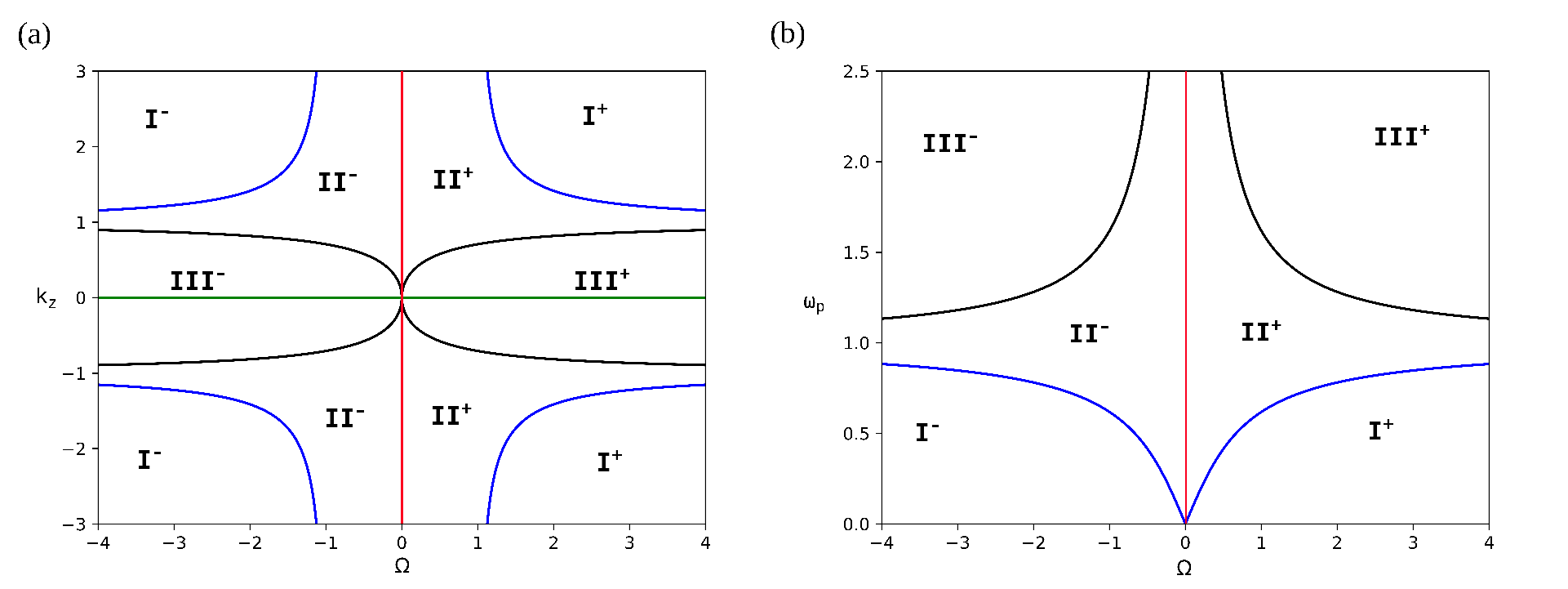}
		\caption{Topological phases plotted in the (a) $\om, k_z$ plane for $\op = 1$ and (b) $\om, \op$ plane for $k_z = 1$. Black lines indicate the $k_z = k_+$ and $\op = \omega_+$ boundaries, blue the $k_z = k_-$ and $\op = \omega_-$ boundaries, and red $\om = 0$ boundary. The green line on the left represents phase IV\sspt{$\pm$} ($k_z = 0$).}\label{fig:1}
\end{figure}

These phase transitions occur as in \cite{qin2023topological} for the values
\begin{equation}\label{eq:omtrans}
    0<\frac{\omega_\pm}{|\om|} = \frac{1}{2}\left(\sqrt{\left(\frac{k_z}{\om}\right)^4 + 4\left(\frac{k_z}{\om}\right)^2} \pm \left(\frac{k_z}{\om}\right)^2\right)
\end{equation}
where $\omega_\pm$ are defined as the transitions $\omega_-=\omega_p=\omega_{L-}$ and $\omega_+=\omega_p=\omega_{R}$. We also note that $\omega_-$ is defined only for under-dense plasma where $\omega_p<\Omega$ while $\omega_+$ is defined for all values of $(\Omega,\omega_p,k_z)$; see Figure \ref{fig:1}.

This allows us to define 6 distinct {\em topological phases}. The phases $\Pha^+$, $\Phb^+$, and $\Phc^+$ are defined for
\[
\begin{aligned}
    \Pha^+ : &\qquad 0<\omega_p<\omega_-<\omega_{L-}
    \\
    \Phb^+ : &\qquad 0<\omega_{L_-} < \omega_- < \omega_p < \omega_+ < \omega_R
    \\
    \Phc^+ :& \qquad 0<\omega_{L_-} < \omega_-  <  \omega_R  < \omega_+< \omega_p < \omega_{L+}.
\end{aligned}
\]
The phase $\Pha^+$ is defined only for under-dense plasmas. The phases $\Pha^-$, $\Phb^-$, and $\Phc^-$ are defined for $\Omega<0$ by the same constraints that $0<\omega_p$ is respectively the smallest, second smallest, or second largest (strictly positive) eigenvalue.  The case $\Omega=0$ is singular and so the above six phases are characterized by different invariants.

\paragraph{Eigenvectors as $k\to\infty$.} In order to apply \eqref{eq:simpChern}, we need to evaluate the behavior as $k\to\infty$ of the above eigenvectors. We do not have explicit expressions for these eigenvectors for arbitrary values of $0<k<\infty$. For $k\to\infty$, we obtain for the four eigenvectors associated to $\omega_1=0$, $\omega_2=\sqrt{\omega_p^2+\Omega^2}$, $\omega_3=k+o(1)$ and $\omega_4=k+o(1)$, up to normalization by:
\begin{equation}\label{eq:psiinf}
\begin{aligned}
    \Psi_1 = \begin{pmatrix}
        \frac{-\hat e_2}{\sqrt{1+\sigma^2}}+i\hat e_3
        \\
        \frac{-\sigma \hat e_1}{\sqrt{1+\sigma^2}}  \\ 0
    \end{pmatrix},\
    \Psi_2 = \begin{pmatrix}
        \frac{-\sigma \hat e_2}{\sqrt{1+\sigma^2}} -i\hat e_1
        \\
         \frac{\hat e_1}{\sqrt{1+\sigma^2}} \\ 0
    \end{pmatrix},\ 
    \Psi_3 = \begin{pmatrix}
     0 \\ \hat e_3 \\ -\hat e_2 
    \end{pmatrix},\
    \Psi_4 = \begin{pmatrix} 0 \\ \hat e_2 \\ \hat e_3
    \end{pmatrix},
\end{aligned}
\end{equation}
where we have defined $\sigma=|\Omega|/\omega_p$ and assumed $k_y = 0$. Expressions for arbitrary $\bk$ can be obtained by applying R$(\theta)^*$ (see Appendix F for details) but since rotations in $\bk$ are unitarily equivalent the above vectors are sufficient to calculate bulk- or bulk-difference invariants. Similar expressions may be obtained for $\Omega<0$ using \eqref{eq:invOmega}.

Using \eqref{eq:simpChern} with $U'(0)=\rR'(0)={\rm Diag}(J,J,J)$ where $J={\rm Diag}( \big(\begin{smallmatrix} 0 &-1\\1 &0 \end{smallmatrix}\big),0) = \hat{e}_z \times$, we deduce the results gathered in Table \ref{tab:1}.
\begin{table}[ht]
    \centering
     \caption{Value of the curvature integrals for the six phases $\Pha^\pm,\Phb^\pm,\Phc^\pm$ and the two phases $\Phd^\pm$ corresponding to $k_z=0$ treated in section \ref{sec:kz} below.}
    \label{tab:1}
    \begin{tabular}{|c|cccc|c|cccc|}
		\hline
		Phase &  $\mathcal{C}_1$&
        $\mathcal{C}_2$& $\mathcal{C}_3$
        &$\mathcal{C}_4$ & 
        Phase &  $\mathcal{C}_1$& $\mathcal{C}_2$ &$\mathcal{C}_3$ & $\mathcal{C}_4$\\
		\hline 
		$\Pha^+$  & 
         $0$ & $\frac{\sigma}{\sqrt{\sigma^2 + 1}}-1$ & $1$ & $-1$ & 
         $\Pha^-$ &
         $0$ & $\frac{-\sigma}{\sqrt{\sigma^2 + 1}}+1$ & $-1$ & $+1$ \\
		\hline
        $\Phb^+$  & 
         $-1$ & $\frac{\sigma}{\sqrt{\sigma^2 + 1}}$ & $1$ & $-1$ & 
         $\Phb^-$ &
         $1$ & $\frac{-\sigma}{\sqrt{\sigma^2 + 1}}$ & $-1$ & $+1$ \\
		\hline     
        $\Phc^+$  & 
         $-1$ & $1+\frac{\sigma}{\sqrt{\sigma^2 + 1}}$ & $0$ & $-1$ & 
         $\Phc^-$ &
         $1$ & $-1-\frac{\sigma}{\sqrt{\sigma^2 + 1}}$ & $0$ & $+1$\\
		\hline 
        $\Phd^+$  & 
         $0$ & $1+\frac{\sigma}{\sqrt{\sigma^2 + 1}}$ & $0$ & $-1$ & 
         $\Phd^-$ &
         $0$ & $-1-\frac{\sigma}{\sqrt{\sigma^2 + 1}}$ & $0$ & $+1$\\
		\hline
	\end{tabular}
   
\end{table}

We observe, as in \cite{qin2023topological}, that these curvature integrals are not always integers. This reflects the fact that $\Rm^2$ is not a compact manifold and that the projectors $\Pi^h_j$ do not have a unique value at $\infty$. Indeed, $\psi(\theta)=\rR(\theta)\psi_0$ in general has explicit dependence in $\theta$.

\subsection{The case $k_z=0$}
\label{sec:kz}
The Hamiltonian family $H(\kp)$ belongs to the same phase ($\Phc^\pm$ depending on the sign of $\Omega$) for all $|k_z|\ll1$; note also the unitary equivalence in \eqref{eq:invkz}. However, the case $k_z=0$ is singular and defines different topological phases since Transverse Magnetic (TM) and Transverse Electric (TE) modes are no longer coupled \cite{silveirinha2015chern,hanson2016notes,marciani2020chiral}. We may indeed introduce in the Fourier variables
\begin{equation}\label{eq:HTMTE}
     H_{TM} = 
    \begin{pmatrix}
        0 & i\Omega & -i\omega_p & 0 & 0 \\
        -i\Omega & 0 & 0 & -i\omega_p & 0 \\
        i\omega_p & 0 & 0 & 0 & -k_y \\
        0 & i\omega_p & 0 & 0 & k_x \\
        0 & 0 & -k_y & k_x & 0 \\
    \end{pmatrix},\qquad H_{TE} = 
    \begin{pmatrix}
        0 & -i\omega_p & 0 & 0\\
        i\omega_p & 0 & k_y & -k_x \\
        0 & k_y & 0 & 0 \\
         0 & -k_x & 0 & 0
    \end{pmatrix},
\end{equation}
where $H_{TM}$ acts on $(v_x, v_y, E_x, E_y, B_z)^T$ while $H_{TE}$ acts on $(v_z, E_z, B_x, B_y)^T$. These Hamiltonians are invariant by a reduction of $\rR(\kp)$ to the appropriate components and we may assume that $\kp=(k,0)$. We find that the eigenvalues of $H_{TE}$ are given by $\check\omega_{1,2,3,4}=(-\sqrt{k^2+\omega_p^2},0,0,\sqrt{k^2+\omega_p^2})$. We verify that all curvature integrals $\mC[\Pi_j]$ associated with their eigenvectors vanish; in particular $\mC_{1,3} = 0$. Due to the fact that the curvature integrals are distinct from phase $\Phc^\pm$, to which $k_z = 0$ would belong otherwise, but still non-trivial, we denote a separate phase $\Phd^\pm$ for the $k_z = 0$ case for $\om > 0$ and $\om < 0$ respectively. In particular note that TE modes acquire a non-trivial topological branch $\mC_1\not=0$ only after coupling with TM modes.

The eigenvalues of $H_{TM}$ are given by $\tilde\omega_{-1,-2}=-\tilde\omega_{1,2}$ and
\[\tilde\omega_{0,1,2}= \Big(0, \frac{1}{2} \left(k^2 + \omega_h^2 - \sqrt{(k^2-\om^2)^2 + 4\op^2\om^2}\right), \frac{1}{2} \left(k^2 + \omega_h^2 + \sqrt{(k^2-\om^2)^2 + 4\op^2\om^2}\right) \Big)\] 
where $\omega_h^2 = 2\op^2 + \om^2$. If $\omega_{0,1,2,3,4}$ denote the nonnegative eigenvalues of the original $9\times9$ system, then we verify that $\omega_{0,1,2,3,4}=(0,\check\omega_1,\tilde\omega_1,\check\omega_2,\tilde\omega_2)$. The numbers $\mC_{2,4}$ in Table \ref{tab:1} are identical in phases $\Phc^\pm$ and $\Phd^\pm$.

\subsection{Reduced models and Dirac Hamiltonians}
\label{sec:reduced}

The first and second spectral bands cross $\omega_1(k = 0)=\omega_2(k = 0)$ at the value $\omega_-$. Similarly the second and third spectral bands cross $\omega_2(k = 0)=\omega_3(k = 0)$ at the value $\omega_+$. Following \cite{qin2023topological} and the reduction introduced in section \ref{sec:BEC}, we derive the corresponding Dirac operators for both $\Omega>0$ and $\Omega<0$.

\paragraph{Crossing near $\omega_-$ frequency.} The regime $|\omega_p-\omega_-|\ll1$ and $|\kp|\ll1$ is characterized by two nearby bands $(\omega_1,\omega_2)$ that are isolated from the rest of the Hamiltonian. Following \cite{qin2023topological}, we project $H$ onto the span of $(\Psi_1(0),\Psi_2(0))$ to obtain the following reduced Hamiltonian $\mathfrak{D}_{ij} = \Psi_i^* H \Psi_j$ given by
\begin{equation}\label{eq:Dirac}
    \mathfrak{D}= \omega_- + \begin{pmatrix} \tilde\omega_p & \frac{ik_x-k_y}\alpha \\ \frac{-ik_x-k_y}{\alpha} & \frac{-2\beta}{\alpha^2} \tilde \omega_p \end{pmatrix} = \Big(\omega_-+(\frac12-\frac{\beta}{\alpha^2})\tilde\omega_p \Big) -\frac{k_y}{\alpha}\sigma_1-\frac{k_x}\alpha\sigma_2 + (\frac12+\frac{\beta}{\alpha^2}) \tilde\omega_p\sigma_3,
\end{equation}
where $\tilde\omega_p=\omega_p-\omega_-$ and $
\alpha^2=4+3(\frac{k_z}\Omega)^2-\frac{k_z}\Omega\sqrt{(\frac{k_z}\Omega)^2+4}$ while $\beta=\frac{k_z}{|\Omega|}(\sqrt{(\frac{k_z}\Omega)^2+4}-\frac{k_z}{|\Omega|}) = 2\omega_-/|\om|$. Here, we assume $\Omega>0$.

The case $\tilde\omega_p<0$ corresponds to phase $\Pha^+$ while $\tilde\omega_p>0$ (not too large) corresponds to phase $\Phb^+$. The reduced interface operator $\mathfrak{D}_I$ with $\tilde\omega_p=\gamma y$ thus acts as a transition from $\mathfrak{D}^S$ in phase $\Pha^+$ to $\mathfrak{D}^N$ in phase $\Phb^+$.

Because $\frac12-\frac\beta{\alpha^2}<\frac12+\frac\beta{\alpha^2}$, we deduce that $\mathfrak{D}_I$ is an {\em elliptic} operator in the sense of \cite{quinn2024approximations,bal2024continuous}.

\paragraph{Crossing near $\omega_-$ frequency for $\Omega<0$.} We now consider the same regime $|\omega_p-\omega_-|\ll1$ and $|\kp|\ll1$ but with now $\Omega<0$. 

The symmetry \eqref{eq:invOmega} implies that $H[-\Omega]=\tilde \Gamma_P H[\Omega]\tilde\Gamma_P$. We observe that likewise projecting $H$ onto  $\tilde\Gamma_P\Psi_1(k = 0)$ and $\tilde\Gamma_P\Psi_2(k = 0)$, yields, for an appropriate choice of phases in $\sigma_1, \sigma_2$: 
\begin{equation}\label{eq:DiracM}
    \mathfrak{D}[-\Omega]= \Big(\omega_-+(\frac12-\frac\beta\alpha)\tilde\omega_p \Big) -\frac{k_y}{\alpha}\sigma_1 + \frac{k_x}\alpha\sigma_2 + (\frac12+\frac\beta\alpha) \tilde\omega_p\sigma_3.
\end{equation}
See Appendix D for details. In other words, $\Omega\to-\Omega$ may be represented as the change of orientation $k_x\to-k_x$.
\paragraph{Crossing near $\omega_+$ frequency.}
A similar calculation involving the projection of $H$ onto the span of $(\Psi_2(0),\Psi_3(0))$ for $\omega_p$ near the band crossing $\omega_+$ reveals the following reduced Hamiltonian:
\begin{equation}\label{eq:DiracP}
   \mathfrak{D}_+= \omega_+ + \begin{pmatrix} \tilde\omega_p & \frac{k_y + ik_x}\alpha \\ \frac{k_y-ik_x}{\alpha} & \frac{2\beta}{\alpha^2} \tilde \omega_p \end{pmatrix} = \Big(\omega_++(\frac12+\frac{\beta}{\alpha^2})\tilde\omega_p \Big) +\frac{k_y}{\alpha}\sigma_1-\frac{k_x}\alpha\sigma_2 + (\frac12-\frac{\beta}{\alpha^2}) \tilde\omega_p\sigma_3,
\end{equation}
where now $\alpha^2=4+3(\frac{k_z}\Omega)^2+\sqrt{(\frac{k_z}\Omega)^4+4\left(\frac{k_z}{\om}\right)^2}$ and $\beta=\frac{k_z}{|\Omega|}(\sqrt{(\frac{k_z}\Omega)^2+4}+\frac{k_z}{|\Omega|}) = 2\omega_+ /|\om|$. Now because $(\frac12 + \frac{\beta}{\alpha^2}) > (\frac12 - \frac{\beta}{\alpha^2})$ we observe that $\omega_+$ is no longer in a spectral gap. The spectral gaps in phases $\Phb^+$ and phases $\Phc^+$ do not overlap, which makes it impossible to have a well defined interface current observable $\sigma_I$ in \eqref{eq:sigmaI} in that setting.
\section{BDI and Edge States: Main results}
\label{sec:main}

Edge states typically appear at interfaces separating insulators in different phases. This requires identifying global spectral gaps that are shared by the two insulators and defining bulk Hamiltonians $H^N$ and $H^S$ with coefficients $(\Omega^N,\omega_p^N)$ and $(\Omega^S,\omega_p^S)$ that place $H^N$ and $H^S$ in different phases according to Figure \ref{fig:1}. 

Since the smallest positive eigenvalue converges to $0$ and the third largest positive eigenvalue tends to infinity as $k\to\infty$,  the only possible spectral gaps are between $\omega_1$ and $\omega_2$ (denoted below by $\ell=1$) or between $\omega_2$ and $\omega_3$ (denoted below by $\ell=2$). Such gaps indeed occur for a range of parameters \cite{fu2021topological}. 

Once such transitions are identified, we wish to assign a topological invariant to the interface Hamiltonian and verify whether the BEC \eqref{eq:BEC} applies. This requires that we be able to construct Chern numbers, either via a one-point compactification of $\Rm^2$ or via a radial compactification of two copies of $\Rm^2$ as recalled in section \ref{sec:invariants}. The construction of such  Chern numbers in \eqref{eq:Chernop} or \eqref{eq:ChernBDI} requires that band projectors satisfy appropriate continuity criteria. We observe from \eqref{eq:psiinf} that one point compactification is not applicable in this case. Even radial compactification and the definition of a BDI requires regularization effect that we will analyze in detail.

Although we treat $k_z$ as a parameter, it is still a parameter of the excitation and not the underlying plasma or material model. Therefore we consider only phase transitions for constant $k_z$. In addition, we emphasize that only continuous variations in $\om, \op$ are considered here. Under these conditions, possible phase transitions and associated spectral gaps are identified as: transition from $\Pha^\pm$ to $\Phb^\pm$ for gaps $\ell=1$ and $\ell = 2$; transition from $\Phb^\pm$ to $\Phc^\pm$ for gap $\ell = 1$; transition from $\Pha^+$ to $\Pha^-$ for gaps $\ell=1$ and $\ell = 2$; transition from $\Phb^+$ to $\Phb^-$ for gaps $\ell = 1$ and $\ell=2$; transition from $\Phd^+$ to $\Phd^-$ for gaps $\ell=1$ and $\ell=2$ for the $5\times5$ TM system. Comparison with Figure \ref{fig:1} verifies that all such transitions can be made by varying $\om, \op$ continuously with constant $k_z$. Additional continuous transitions with global spectral gaps are possible, but only as a superposition of the above transitions- e.g. for $\ell = 1$, a gapped $\Phc^-$ to $\Phc^+$ transition can be made continuously with $k_z$ constant only as a superposition of transitions $\Phc^- \rightarrow \Phb^-\rightarrow \Phb^+\rightarrow\Phc^+$. Therefore analyzing the above transitions fully characterizes any continuous topological phase transitions when $k_z$ is held constant.

Numerical diagonalization of the interface Hamiltonian $H_I$ via finite differences is used to verify the BEC in each of the above cases. Of particular note is the fact that some branches of spectrum appear to be discontinuous. This is a product of our choice of periodic boundary conditions, which necessarily produce edge modes both at the transition of interest at $y = 0$ and at a spurious phase transition at the edge of our domain $y = L$ introduced by periodic boundary conditions. We choose to eliminate edge modes which appear at this spurious boundary for clarity but occasionally the same branch of spectrum will produce edge modes at $y = 0$ for $k_x > 0$ ($k_x <0$) and at $y = L$ for $k_x < 0$ ($k_x >0)$, giving the appearance of a discontinuous branch of spectrum. Symmetry arguments and inclusion of spurious modes shows that these branches do not cross the band gap and thus do not contribute to $\sigma_I$, see Appendix E for details.

We now analyze these gaps and their topological properties, selecting $\pm$ above as being equal to $+$ for concreteness. The predicted number of edge states and the validity of the BEC \eqref{eq:BEC} is illustrated in each case by numerical simulations of the spectral decomposition of the interface Hamiltonian.

\paragraph{BDI $\fC[\Pha^+_\ell,\Phb^+_\ell]$.} 
Consider the transition from $\Pha^+$ to $\Phb^+$. Thanks to the reduced model \eqref{eq:Dirac} with $\tilde\op=\gamma y$ with $\gamma>0$ the Dirac operator is in phase $\Pha^+$ for $y<0$ and in phase $\Phb^+$ for $y>0$. In other words, $\omega_p^S<\omega_-(\Omega,k_z)<\omega_{L-}^S$ in phase $\Pha^+$ while $\omega_{L-}^N<\omega_-(\Omega,k_z)<\omega_p^N$ in phase $\Phb^+$.

We now wish to construct a BDI by radial compactification as one-point compactification cannot hold without regularizing the problem. Thanks to the invariance \eqref{eq:invrot}, it is sufficient to verify the gluing condition \eqref{eq:gluing} at $\theta=0$ for $\ell=1$ (and we equivalently verify gluing for $\sum_{j\geq \ell} \Pi_j^h$).
However, we observe that while $\Psi_{3,4}$ are the same in phases $N$ and $S$, this is not the case for $\Psi_2$ in \eqref{eq:psiinf} since $\sigma$ depends on $\omega_p$ and $\omega_p^N\not=\omega_p^S$ as a necessary condition to transition from $\Pha^+$ to $\Phb^+$ which sustains a global band gap.

\begin{figure}[ht]
			\centering
			\includegraphics[width=\textwidth]{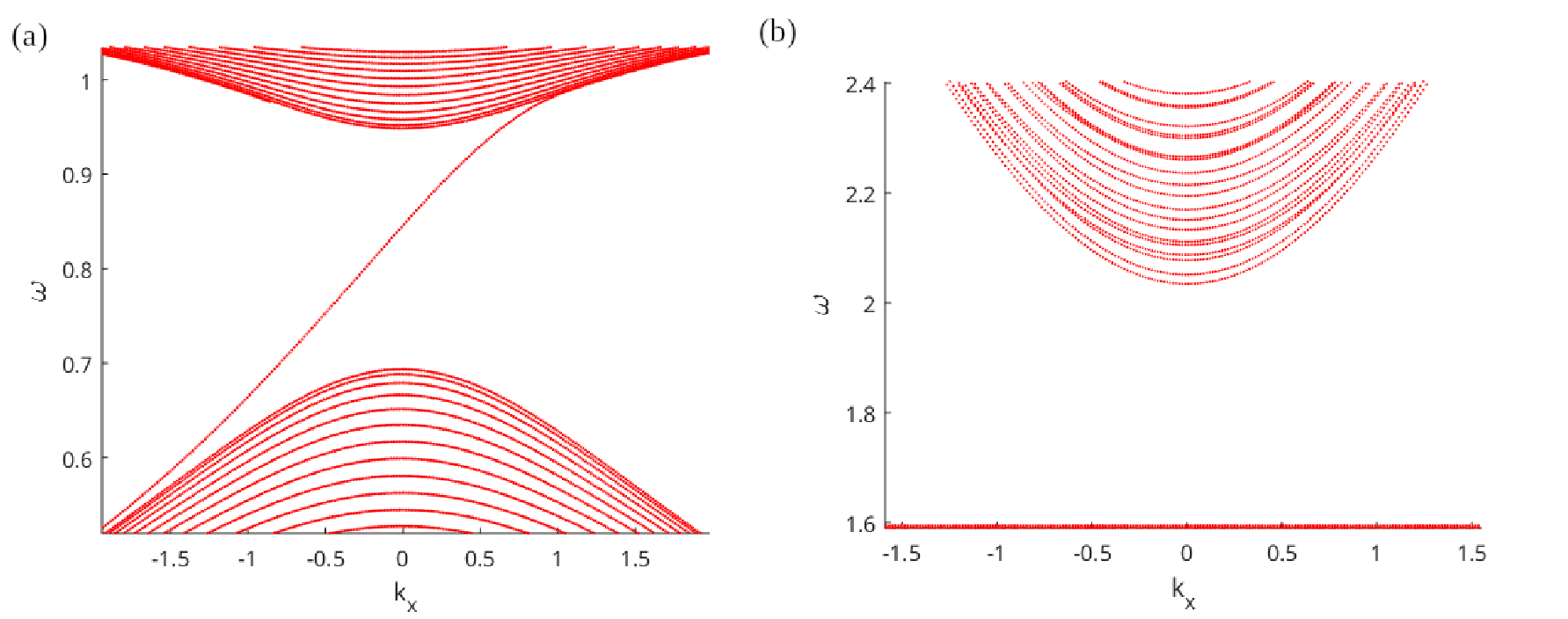}
		\caption{Numerically calculated spectrum with the parameters $(k_z, \om) = (2, 1)$ for the transition from phase I\sspt{+} to II\sspt{+} ($\op$ varying from $0.5\omega_-$ to $1.5\omega_-$ with $\omega_- =$ 0.8284 s\sspt{-1}). (a) the spectral gap between bands 1 and 2, which illustrates one topologically protected edge state. (b) the spectral gap between bands 2 and 3, which has no edge states as predicted. 
        }
        \label{fig:I+II+}
\end{figure}

For this reason, we assume that $\Omega=\Omega(k)$ and $\omega_p=\omega_p(k)$ in the definition \eqref{eq:HFourier}. We observe that $\Psi_2^N=\Psi_2^S$ (up to a multiplicative phase in $U(1)$) provided that $\bar\sigma=\lim_{k\to\infty} |\Omega(k)|/\omega_p(k)$ exists (and is independent of the values of $(\Omega(0),\omega_p(0))$). In this case, we obtain from Table \ref{tab:1} a well-defined invariant for the gap $\ell=1$ given by \cite{qin2023topological}
\begin{equation}\label{eq:transItoII}
    \fC[\Pha_1^+,\Phb_1^+] = (\frac{\bar\sigma}{\sqrt{1+\bar\sigma^2}} +1-1)-(\frac{\bar\sigma}{\sqrt{1+\bar\sigma^2}}-1 +1-1)=1.
\end{equation}
The above regularization generalizes the regularization models used in \cite{qin2023topological,silveirinha2015chern}, which consists in assuming that $\omega_p(k)\to0$ as $k\to\infty$. This edge mode was studied in \cite{parker2020topological, fu2021topological, fu2022dispersion, qin2023topological} and termed the Topological Cyclotron-Langmuir Wave (TCLW). Importantly \cite{parker2020topological} conducted numerical studies using a cylindrical geometry and parameters achievable in operating plasma devices, showing that this mode may be accessible experimentally using devices such as the Large Plasma Device. The predicted number of edge modes is confirmed in the simulation shown in Figure \ref{fig:I+II+}.

The above invariant is consistent with the BDI associated to the reduced Dirac operator in \eqref{eq:Dirac}, where \cite{quinn2024approximations}
\[  \fC[\mathfrak{D}^S,\mathfrak{D}^N] = 1.\]

We can similarly calculate that for $\ell = 2$:
\begin{equation}\label{eq:transItoII2}
    \fC[\Pha_2^+,\Phb_2^+] = (1-1) - (-1 +1) = 0,
\end{equation}
which is also verified numerically in Figure 3. Since $\Psi_3^N, \Psi_4^N$ are the same as $\Psi_3^S, \Psi_4^S$ \eqref{eq:gluing} holds without need for regularization in this case.
\paragraph{BDI $\fC[\Phb^+_\ell,\Phc^+_\ell]$.} 
As we discussed in Section 3.5, there is no fully gapped transition from phase $\Phb^+$ to $\Phc^+$ for $\ell = 2$. However, the frequency branches 2 and 3 are still separated even if there is no global gap and the construction of a BDI is in fact independent of the presence of a global gap. Using the same regularization procedure as defined above, we would obtain a BDI equal to $\fC[\Phb^+_2,\Phc^+_2] = (0-1)-(1-1)=-1$ involving the gluing of $\Psi_{3,4}$, which is guaranteed without the need to regularize the problem. Due to the fact that $\sigma_I$ is ill-defined as discussed in Section 3.5, however, we cannot assess the BEC with respect to this BDI with numerical spectral calculations.

\begin{figure}[h!]
        \centering
        \includegraphics[width = .5\textwidth]{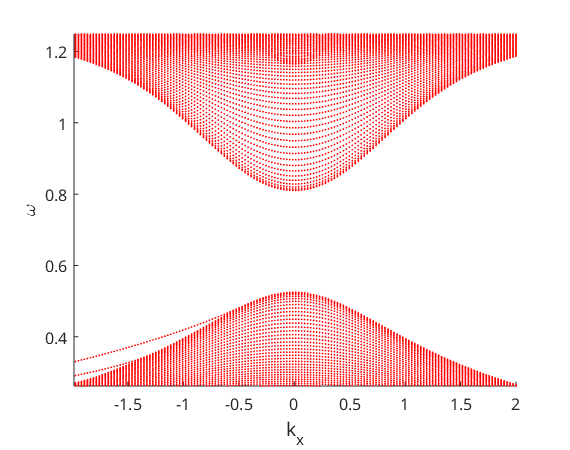}
        \caption{Numerically calculated spectrum with parameters $(k_z, \om) = (1, 1)$ for the transition from II\sspt{+} to III\sspt{+} ($\op$ varying from $0.5\omega_+$ to $1.5\omega_+$ with $\omega_+ = 1.33$ s\sspt{-1}).}
        \label{fig:II_III+}
    \end{figure}

A global band gap is possible for $\ell = 1$. Using the same regularization as in the previous paragraph, $\bar\sigma=\lim_{k\to\infty}|\om(k)|/\op(k)$, the BDI is well-defined for this gap and we compute:
\[
\fC[\Phb_1^+,\Phc_1^+] = (1+\frac{\bar{\sigma}}{\sqrt{\bar{\sigma}^2 + 1}}-1) - (\frac{\bar{\sigma}}{\sqrt{\bar{\sigma}^2 + 1}} + 1-1) = 0
\]
This result is verified numerically in Figure \ref{fig:II_III+}.

\paragraph{BDI $\fC[\Phb_\ell^-,\Phb_\ell^+]$.} We now consider the transition for $\ell=1$ between phases $\Phb^+$ and $\Phb^-$ assuming $\Omega^S<0$ and $\Omega^N>0$ in the construction of the interface Hamiltonian $H_I$. 
We observe that $\Gamma_P(0)\Gamma_\Omega={\rm Diag}(1,-1,1,1,-1,1,-1,1,-1)$ and we therefore need to glue together $\Psi_2$ and $\Gamma_P(0)\Gamma_\Omega\Psi_2$ in \eqref{eq:psiinf} (modulo a phase in $U(1)$) to define a BDI. 

\begin{figure}[ht]
        \centering			
        \includegraphics[width=1\textwidth]{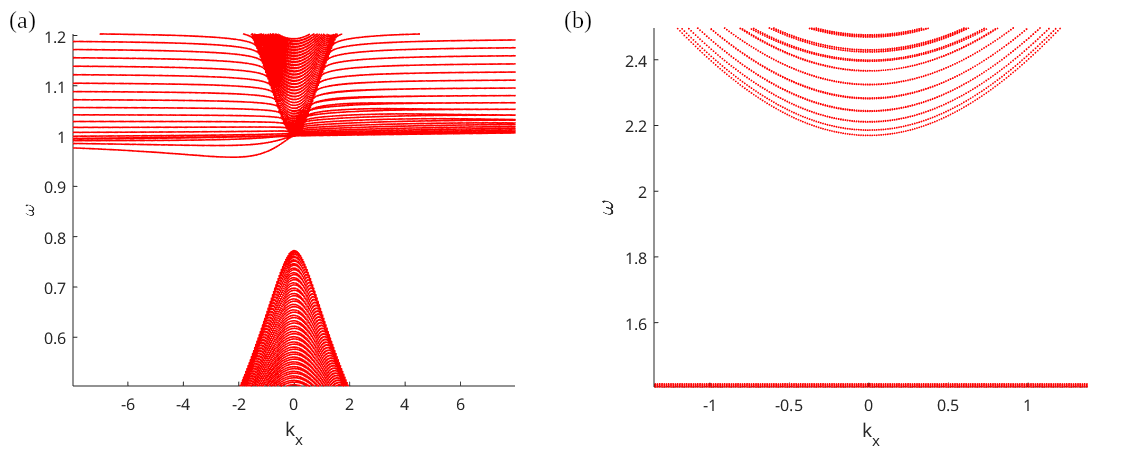}
		\caption{Numerically calculated spectrum with the parameters $(k_z, \op) = (2, 1)$ for the transition from phase II\sspt{-} to II\sspt{+} with $\om$ varying from -0.75 to 0.75 s\sspt{-1}. (a) shows the spectral gap between bands 1 and 2 and (b) the gap between bands 2 and 3. No edge states are present as predicted by BDI's. 
        }
        \label{fig:II-II+}
\end{figure}

This is possible only when the term involving $\hat e_2$ vanishes as it transforms with a different sign from the other components. As a result, we need a regularization implying that $\bar\sigma=0$ in the limit $k\to\infty$. We thus propose the regularization 
\begin{equation}\label{eq:regulOm}
    \Omega(k) = (1+\eta|k|^2)^{-\frac12} \Omega(0),
\end{equation}
with $\eta>0$ arbitrarily small. This allows one to obtain 
\begin{equation}\label{eq:BDIII}
    \fC[\Phb_1^-,\Phb_1^+] = (0+1-1)-(0-1+1) =0.
\end{equation}
This predicts the absence of topologically protected asymmetry separating phases $\Phb^-$ and $\Phb^+$, which we verify numerically in Figure \ref{fig:II-II+}.

In contrast, a regularization imposing $\omega_p(k)\to0$ as $k\to\infty$, as in, e.g., \cite{silveirinha2015chern,qin2023topological}, leads to $\bar\sigma=\infty$ and a predicted an invariant equal to $(1+1-1)-(-1-1+1)=2$. Note that the latter invariant, which is simply obtained by summing curvature integrals, is not a BDI since the projectors associated to $\Psi_2^N$ and $\Psi_2^S$ cannot be glued together continuously in a radial compactification. The numerical simulations in Figure \ref{fig:II-II+} confirm that the predicted number of protected edge modes $2$ is incorrect while $\fC[\Phb^-_1,\Phb^+_1]=0$ is correct. Aside from the fact that \eqref{eq:regulOm} produces well-defined BDI's which agree with numerical results, we provide some physical justification for \eqref{eq:regulOm} in Appendix C as a high-wave number limit of a hydrodynamic model which allows for time-dependent perturbations in electron density. 

For $\ell = 2$ we again can rely on the fact that $\Psi_{3, 4}$ are the same in phases $N$ and $S$, so without regularization the BDI is well defined and:
\begin{equation}\label{eq:BDIII2}
    \fC[\Phb_2^-,\Phb_2^+] = (1-1)-(-1 +1)=0.
\end{equation}

Although it is clear that well defined BDI's predict 0 edge modes in this transition, a more heuristic argument can also be made for why this must be the case. Calculating the eigenvalues of \eqref{eq:HFourier} with $\om = 0$ is straightforward and produces positive eigenvalues $(\omega_1, \omega_2, \omega_3, \omega_4) = (0, \omega_p, \sqrt{\op^2 + |\kop|^2}, \sqrt{\op^2 + |\kop|^2})$. In phase $\Phb^\pm$ we have that $k_z \ne 0\Rightarrow |\kop|^2 > 0$, therefore in the transition $\Phb^-\rightarrow \Phb^+$ band 2 in fact does not cross any band. Hence band 2 cannot (at least in the continuous sense) exchange modes with another band, and all BDI's involving band 2 should intuitively be 0.

\paragraph{BDI $\fC[\Phd^-_\ell,\Phd^+_\ell]$.}

This corresponds to the case $k_z=0$ and the analysis of the $5\times5$ system for TM modes. Focusing on TM modes, we observe the presence of two global gaps; a first one between $\tilde\omega_0=0$ and $\tilde\omega_1=\omega_2$ and a second one between $\tilde\omega_1=\omega_2$ and $\tilde\omega_2=\omega_4$. The second gap is actually filled by TE modes corresponding to $\omega_3$. We thus have a new gap not present in the $9\times9$ system, whose topological properties we now analyze. 

\begin{figure}[ht]
			\centering
			\includegraphics[width=\textwidth]{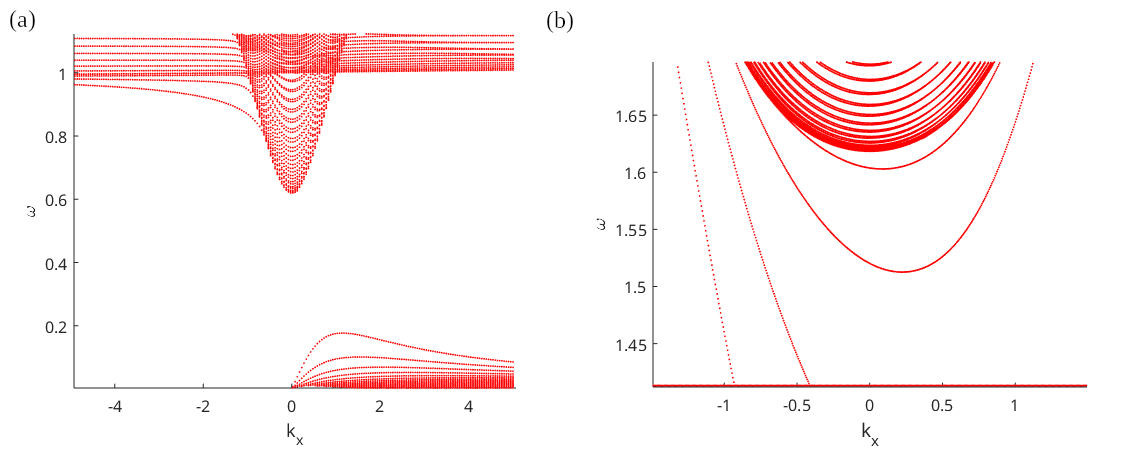}
		\caption{Numerically calculated spectrum for transition $\Phd^-\to \Phd^+$ with parameters $(k_z, \op) = (0, 1)$ where $\om$ varies from -1 to 1 s\sspt{-1}. (a) shows the gap between transverse magnetic modes 1 and 2, showing no edge modes. (b) shows the spectral gap between transverse magnetic modes 2 and 3, demonstrating 2 topologically protected edge modes as predicted.   Both are consistent with prediction of BDI's.
        }
        \label{fig:IV}
\end{figure}

As in the transition from $\Phb^-$ to $\Phb^+$, we assume the regularization \eqref{eq:regulOm}. We thus obtain the two phase transitions (considering terms $\mC_{2,4}$ in Table \ref{tab:1} while discarding $\mC_{1,3}$, which are trivial anyway) corresponding to TM modes:
\begin{equation}\label{eq:BDIIV}
    \fC[\Phd^-_1,\Phd^+_1] = (+1-1)-(-1+1) =0,\quad  \fC[\Phd^-_2,\Phd^+_2] = (-1)-(+1) =-2.
\end{equation}
One of the transitions is therefore topologically trivial while the second one predicts the presence of two edge modes.  We see in the numerical simulations of Figure \ref{fig:IV} that two edge modes are indeed present in the gap $[\tilde\omega_1,\tilde\omega_2]$ while none are present in the gap $[0,\tilde\omega_1]$ for an interface Hamiltonian $H_I$ where $\Omega(y)$ transitions from $-\Omega_0<0$ for $H^S$ in phase $\Phd^-$ to $+\Omega_0>0$ for $H^N$ in phase $\Phd^+$ (strictly speaking these absolute phases are not described by an invariant; only the phase difference is). Notably, this transition may be accessible in the THz range in InSb \cite{wang2019photonic}, whose electronic structure behaves as cold plasma, due to the fact that varying magnetic fields in photonic settings is much more realistic than in large-scale plasma devices.

Note that a regularization based on $\omega_p(k)\to0$ as $k\to\infty$ (with $\sigma=\infty$ in Table \ref{tab:1}) would predict phase transitions in these two spectral gaps given respectively by an incorrect $(+2-1)-(-2+1)=2$ for $\ell = 1$ and a correct $(-1)-(+1)=-2$ for $\ell = 2$. We reiterate that these transitions, while described by integers, are not BDI as the gluing conditions \eqref{eq:gluing} are not met.

It should be noted that many recent results \cite{gangaraj2018coupled,hassani2020physical,buddhiraju2020absence,hassani2019truly} have shown that there may be violations of the BEC in the transition $[\Phd^-_2, \Phd^+_2]$. Many of these arguments are based on physical limitations of the particular photonic/cold plasma model and non-local effects which are not considered here \cite{hassani2020physical,buddhiraju2020absence,hassani2019truly}. However, in \cite{gangaraj2018coupled} it was shown numerically and by Green's function analysis that only one topologically protected edge mode exists in the gapped transition $[\Phd^-_2, \Phd^+_2]$ although 2 were still predicted. While it would seem that our results differ, a key difference in our assumptions is the insistence of continuity of topological phase transitions, e.g. continuity of $\op(y), \om(y)$. It was shown in \cite{bal2024topological} that in a $3\times 3$ PDE model of shallow water equations which govern equatorial waves the BEC holds only when the parameters of the system are continuous. We expect that a similar result holds here and the effect of boundary regularity on BEC for photonic systems is the subject of future work. In addition, our analysis shows that these edge modes are not robust to perturbations in $k_z$ around 0 so that even small deviations from $k_z = 0$ will couple topologically protected TM modes to bulk TE modes which fill the $\ell = 2$ TM spectral gap and allow dispersion into the bulk.

\paragraph{BDI $\fC[\Pha^-,\Pha^+]$.} We note that the phases $\Pha^-$ and $\Pha^+$ are not connected in the diagram in Figure \ref{fig:1} except through the single point $(\om, \op) = (0, 0)$. Still, we may envision a transition between these phases since they share a common gap for $\ell=1$ and $\ell = 2$. We verify that eigenvectors may be glued together under the regularization assumption \eqref{eq:regulOm}. We then obtain for the gap $\ell=1$ the following invariant
\begin{equation}\label{eq:BDII-I+}
    \fC[\Pha_1^-,\Pha_1^+] = (-1+1-1)-(1-1+1)=-2.
\end{equation}

Similarly, for $\ell = 2$ we may verify without need for regularization that:
\begin{equation}\label{eq:BDII-I+2}
    \fC[\Pha_2^-, \Pha_2^+] = (1-1) -(-1+1) = 0
\end{equation}

Figure 7 illustrates the numerically calculated spectrum of the interface Hamiltonian for this transition. While the BEC for the $\Pha_2^-\to \Pha_2^+$ case is verified, we observe that no topologically protected edge states appear numerically for the $\Pha_1^- \to \Pha_1^+ $ case, contrary to our BDI, and in fact the band gap which is expected in this case is filled by a continuum of flat bands. This represents the only case where the BEC fails for well-defined BDI, and the specific mechanism of failure is analyzed extensively in Section \ref{sec:disc}.

\begin{figure}[ht]
			\centering
			\includegraphics[width=\textwidth]{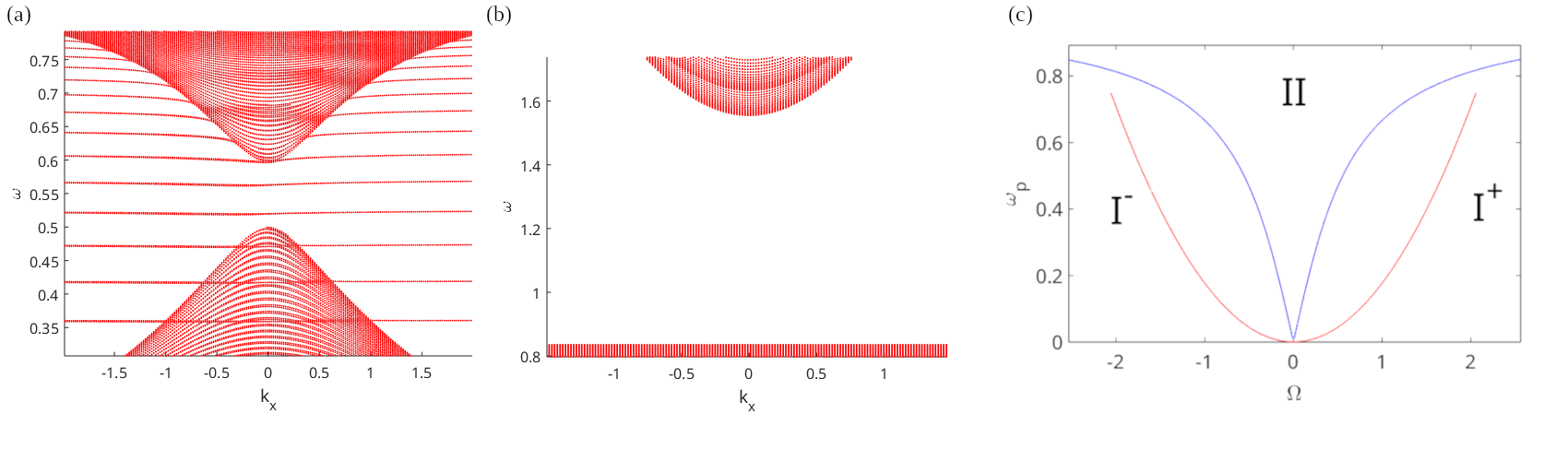}
		\caption{Numerically calculated spectrum for the I\sspt{-} to I\sspt{+} transition. (a) is the numerically calculated spectrum for $\ell = 1$ using the profile in (c), showing a gap in bulk spectra, but the gap being filled by continuum modes as described in Section \ref{sec:disc}. (b) shows the respective $\ell = 2$ spectral gap, showing no edge modes. (c) shows $\op$ as a function of $\om$ for the phase transition considered plotted in red, which necessarily must pass through $(\op, \om) = (0, 0)$ for such a transition to occur. Plotted in blue is the $\omega_-$ boundary between phase $\Pha$ and phase $\Phb$ as in Figure \ref{fig:1}. }
        \label{fig:ImIp}
\end{figure}

\section{Limitations of BEC}
\label{sec:disc}

Table \ref{tab:2} summarizes the BDI's we obtained in Section 4 for each possible phase transition. To obtain these well-defined BDI's two separate high-wavenumber regularizations are possible. First, for the transitions $\Pha^{\pm} \rightarrow \Phb^{\pm}$ and $\Phb^{\pm} \rightarrow \Phc^{\pm}$ (or any combination of phases in which $\om$ does not change sign) any regularization for which we obtain a unique limit $\lim_{k\rightarrow \infty} |\om(k)|/\op(k) = \bar{\sigma}$ is valid. However, for any transition which includes a change in sign of $\om$, we must impose a regularization which obeys the stricter condition $\lim_{k\rightarrow \infty} |\om(k)|/\op(k) = 0$. We chose \eqref{eq:regulOm} for simplicity and give some heuristics as to why this might be valid in Appendix C. However, this is not the only valid choice and we emphasize that any high-wavenumber regularization which produces the above limit will produce the same well-defined BDI's as shown in Table \ref{tab:2}.

\begin{table}[ht]
    \centering
     \caption{Summary of BDI's calculated for each possible phase transition.}
    \label{tab:2}
     \begin{tabular}{|c|c|l|}
        \hline
        Phase Transition & BDI $(\ell = 1, \ell = 2)$\\
        \hline
        $\Pha^{\pm} \rightarrow \Phb^{\pm}$ & $(\pm 1, 0)$\\
        \hline
        $\Phb^{\pm} \rightarrow \Phc^{\pm}$ & $(0, \mp 1)$\\
        \hline
        $\Pha^{-} \rightarrow \Pha^{+}$ & $(-2, 0)$\\
        \hline
        $\Phb^{-}\rightarrow \Phb^{+}$ & $(0, 0)$\\
        \hline
        $\Phd^- \rightarrow \Phd^+$ & $(0, -2)$\\
        \hline
    \end{tabular}
   
\end{table}

As to the validity of the BEC using these results, one can readily verify by comparing Table \ref{tab:2} with Figures 3-7 that the BEC holds in all cases except two: $\Phb^{\pm}_2 \rightarrow \Phc^{\pm}_2$ and $\Pha^{-}_1 \rightarrow \Pha^{+}_1$. In the former case, we have seen through \eqref{eq:DiracP} that, even locally, no common band gap exists in the $\ell = 2$ case. Therefore, although the BDI is well-defined, $\sigma_I$ is not well defined due to the absence of a common band gap. The $\fC[\Pha^{-}_1 ,\Pha^{+}_1]$ case, on the other hand, merits further investigation.

When considering a 2x2 reduced model of the $\Pha_1^- \to\Pha_1^+$ transition, we will now show that this transition is in fact non-elliptic, and $\sigma_I$ in this case is similarly ill-defined. Therefore, $\fC[\Pha^{-}_1 ,\Pha^{+}_1]$, even as it takes the form of a BDI, should not be trusted. To demonstrate this, we use the reduced Dirac models introduced in \eqref{eq:Dirac} and \eqref{eq:DiracM} to describe the $\ell = 1$ band gap for the $\Pha_1^- \to \Pha_1^+$ transition. 
For both $\om > 0$ and $\om < 0$, we assume that $\tilde\omega_p<0$ so as to ensure that the system remains in Phase $\Pha$. The transition from $\Pha^-$ to $\Pha^+$ at the level of reduced models therefore corresponds to a transition from $\mathfrak{D}^S=\mathfrak{D}[-\Omega]$ to $\mathfrak{D}^N=\mathfrak{D}[\Omega]$ as defined by \eqref{eq:Dirac} and \eqref{eq:DiracM}. By comparing \eqref{eq:Dirac} and \eqref{eq:DiracM} with \eqref{eq:Diracgal} we can see that this transition may be modeled by an anisotropic Fermi velocity $v_x(y)$ transitioning from $v_x(y)<0$ for $y\leq-1$ to $v_x(y)>0$ for $y\geq1$ while $v_y(y)$ remains the same sign for both $y> 0$ and $y< 0$. This reduced model is analyzed in detail below. Analysis of these singular Dirac-type Hamiltonians is novel and may be of interest when studying topological phase transitions beyond the cold plasma model.

\paragraph{Singular Hamiltonian I.}
First, consider a slight modification of \eqref{eq:Diracgal} with $\eta=0$:
\[
 \tilde \fD_I = e^{-i\theta(y)\sigma_2} H_I  e^{i\theta(y)\sigma_2},\quad 
 \fD_I=v_x(y)D_x\sigma_1+D_y\sigma_2+m\sigma_3,
\]
where $v_x(y)=1$ for $y\geq1$ and $v_x(y)=-1$ for $y\leq-1$ and where $\theta(y)$ is a smooth function such that $\theta(y)=0$ for $y\geq1$ and $\theta(y)=\frac\pi2$ for $y\leq -1$. We observe that $e^{\pm i\frac\pi2\sigma_2}=\pm i\sigma_2$. Thus 
\[ \tilde \fD^N=D_x\sigma_1+D_y\sigma_2+m\sigma_3, \qquad 
\tilde \fD^S=D_x\sigma_1+D_y\sigma_2-m\sigma_3.\]
No bulk phases may be obtained for the unregularized $\tilde H^h$ for $h\in\{N,S\}$ but a BDI may be defined \cite{bal2024continuous,quinn2024approximations} as
\[
\fC[\tilde \Pi_1^S,\tilde \Pi_1^N] = \BDI = -1.
\]
The presence of $\theta(y)\not=0$ is there to ensure that $\tilde \Pi_1^S$ and $\tilde \Pi_1^N$ satisfy the gluing condition \eqref{eq:gluing}, which would not hold for $\theta\equiv0$. Now, $\tilde \fD_I$ is unitarily equivalent to $\fD_I$, which in Fourier variables is $\hat \fD_I=v_x(y)\xi \sigma_1 + D_y\sigma_2+m\sigma_3$. This operator
is however gapped in $(-m,m)$. We are thus in a situation where
\[ 
2\pi \sigma_I[\tilde \fD_I]=0,\qquad \fC[\tilde \Pi_1^S,\tilde \Pi_1^N] =\BDI=-1
\]
in complete contradiction of the BEC. The change of sign of $v_x(y)$ in a leading term of the Dirac operator is a strong violation of ellipticity. 

%
\paragraph{Singular Hamiltonian II.} 
Next, consider the Hamiltonian \eqref{eq:Diracgal} with $\eta=0$ and with $v_x(y)=y$ while $v_y(y)=m(y)=|y|$. As discussed above, this model corresponds to the reduced two-band model for the $\Pha_1^- \to \Pha_1^+$ transition. These functions could be assumed smoother without affecting the results we obtain. This highly singular Hamiltonian has the same bulk properties $\BDI=-1$ as above.
We now show that the whole line $\Rm$ belongs to the essential spectrum of $\fD_I$ so that $\sigma_I[\fD_I]$ is not defined as $i[\fD_I,P]\varphi'(\fD_I)$ is not trace-class and \eqref{eq:BEC} cannot possibly hold.

Consider the standard change of variables $\Rm_+\ni y=e^z$ for $z\in\Rm$ so that $dy=ydz$. Using the corresponding  map $\varphi(z)=y^{\frac12}\psi(z)$, we observe that $\|\psi\|=\|\varphi\|$  so that the map is unitary and that moreover $\fD_I\psi=E\psi$ is equivalent to
\[
 \fD_z \varphi(z) =( e^z\xi \sigma_1 + D_z \sigma_2 + e^z \sigma_3)\varphi(z) = E \varphi(z). 
\]
The above shows that $\fD_I$ and $\fD_z$ are unitarily equivalent. We now observe that $\Rm$ belongs to the spectrum of $\fD_z$ for each $\xi\in\Rm$. Indeed, let $\chi(z)\in C^\infty_c(\Rm)$ be such that $\int_{\Rm} \chi^2(z)dz=1$ and $\int_{\Rm} (\chi')^2(z)dz<\infty$ and such that $\chi(z)=0$ for $z\geq-1$ and $z\leq -3$. We next define the sequence
\[
  \varphi_N(z) = \frac{1}{\sqrt N} \chi(\frac zN) e^{-iEz} \frac1{\sqrt2} \begin{pmatrix} 1 \\ i \end{pmatrix}.
\]
We observe that 
\[
  (\fD_z-E) \varphi_N(z) =  e^{-N} r_N(z) + \frac{1}{N} \chi'(\frac zN) s_N(z)
\]
with $r_N$ square integrable and $s_N$ bounded.  This implies that $\varphi_N$ is a Weyl sequence normalized in $L^2(\Rm)$ and without accumulation point as $N\to\infty$.  This further implies that 
\[
  \psi_N(z) = y^{-\frac12} \varphi_N(\ln y)
\]
is also a Weyl sequence, also without accumulation points as supports are disjoint for large, sufficiently different, values of $N$. Standard results in spectral theory \cite[Chapter 2]{teschl2014mathematical} show that $\Rm$ is in the essential spectrum  of $H$ with Weyl sequence independent of $\xi$, which heuristically looks like a continuum of flat bands. Clearly, for $\varphi'$ supported near the origin, then $i[H,P]\varphi'(H)$ is not trace-class so that $\sigma_I$ cannot be defined. Note that these functions can be extended by $0$ for $y<0$ while another set of Weyl sequences can be constructed with support in $y<0$.

This gives an example where $\BDI=-1$ does not predict the edge mode structure. In fact, we have a continuum of modes that decay like $|y|^{-\frac12}$ and hence while localized near the interface $y=0$ cannot really be considered as edge modes. The linear vanishing of the Fermi velocity creates singular operators for which the BEC is incorrect. This highly singular Hamiltonian essentially ensures that the half spaces $y>0$ and $y<0$ are independent.  The bulk-edge correspondence, implying that edge states are independent of the transition between the bulk phases,  cannot possibly hold.

Therefore, the BDI $\fC[\Pha^{-}_2 ,\Pha^{+}_1]$, even as a well-defined BDI, should not be trusted to provide information on the interface current observable and the number of edge modes, given that the reduced model is non-elliptic and $\sigma_I[\fD_I]$ is in fact ill-defined as in the $\Phb_2^+\to\Phc_2^+$ case. This is confirmed by numerical simulations in Figure \ref{fig:ImIp}(a), where transitions for the $9\times9$ system $H_I$ from negative to positive values of $\Omega$ yield no discernible edge state, but rather displays a continuum of flat bands as predicted above.

\section{Conclusion}
For a magnetically biased coupled light-matter Hamiltonian that finds applications in cold plasmas and macroscopic descriptions of photonic materials, we compute integrals of Berry curvature for the eight possible topological phases the material can take. We identified common spectral gaps among topological phases, essentially recovering earlier work in \cite{parker2020topological,silveirinha2016bulk,hanson2016notes,qin2023topological}. Without regularization, the integrals of Berry curvature are not integer-valued and do not represent stable topological invariants. A standard regularzation technique, introduced in \cite{silveirinha2015chern}, restores integer values to integrals of Berry curvature and allowed for the correct prediction of topologically protected edge states in two important cases \cite{fu2021topological, silveirinha2015chern,parker2020topological}.

Our main claim is that having integer-valued curvature integrals is not sufficient to predict the number of edge states at interfaces separating different bulk phases. Rather, regularization should aim to construct bulk difference invariants, which give a correct prediction of the bulk edge correspondence for elliptic operators. We constructed such BDI's using a novel regularization technique which has justification as the high-wave number limit of hydrodynamic considerations. Numerical spectral calculations verify that these BDI's predict the correct number of topologically protected edge states in all cases except one, in which the Hamiltonian is singular and cannot be expected to obey the BEC.

\section*{Acknowledgments} 
This work was funded in part by NSF grant DMS-230641.


%
%

\appendix
\section*{Appendices:}

\section{Invariants for continuous operators}
\label{sec:PDOinvariants}
We recall the definition of invariants for continuous operators following the review in \cite{bal2024continuous}. Any continuous differential operator may be written as 
\begin{equation}\label{eq:weylquantization}
  H_If(x) = (\ow \alpha) f(x)  = \dint_{\Rm^{2d}} \dfrac{e^{i(x-y) \cdot\xi}}{(2\pi)^d}  \alpha(\frac{x+y}2,\xi) f(y)  d\xi dy.
\end{equation}
For the (possibly regularized) Dirac operator in dimension $d=2$ with $y$-dependent coefficients, we have for instance
\[
\alpha(y,\xi,\zeta) = v_x(y) \xi\sigma_1+\zeta\sigma_2 + (m(y)+\eta(\xi^2+\zeta^2)) \sigma_3.
\]
Associated to $H_I$ is an operator
\[  F = H_I-ix = \ow(\alpha-ix).\]
Under general conditions, $F$ is a Fredholm operator, whose index ${\rm Index F}={\rm dim Ker}F - {\rm dim Ker} F^*$ is given by the formula
\begin{equation}\label{eq:FH}
    {\rm Index F}= {\rm F}_2[a] := \frac{1}{24\pi^2} \dint_{ \Sm^4_R }  {\rm tr}  (a^{-1}da)^{\wedge\,3},
\end{equation}
where $\Sm^4_R$ is the 4-sphere in phase space of radius $R$. Here, $R$ is chosen large enough that $a^{-1}$ is defined on it. For the Dirac operator, where the singularities occur at $(x,y,\xi,\zeta)=0$, any $R>0$ would do. This integral may be estimated in various ways \cite{bal2024continuous,quinn2024approximations} and gives the result ${\rm F}_2[a]=-\sgn{v_x}$ for $m(y)=y$ or $m(y)=\sgn{y}$ for instance and $v_x\not=0$ constant. For $v_x(y)=\arctan y$ and $\eta>0$, then we find ${\rm F}_2[a]=-2$. 

Under general hypotheses satisfied by Dirac operators and more generally elliptic operators, we have that 
\[ {\rm F}_2[a] :=  \dint_{ \mC_{L} }  {\rm tr}  (a^{-1}da)^{\wedge\,3},
\]
where $\mC_{L} =\{y=L\} \cup \{y=-L\}$. This implies that ${\rm F}_2[a]$ is a BDI since it depends on the coefficients of $H_I=\ow\alpha$ only at $y=\pm R$ in the N and S hemispheres. Moreover, some calculations \cite{bernevig2013topological,bal2024continuous} show that 
\[ {\rm F}_2[a] = \BDI = \fC[\Pi^S,\Pi^N]. \]
For elliptic operators, we further obtain that the BEC \eqref{eq:BEC} holds. In this context, ellipticity means that $\alpha(y,\xi,\zeta)$ has singular values that grow at least linearly as $(\xi,\zeta)\to\infty$. Other ellipticity conditions are proposed in \cite{bal2022topological,quinn2024approximations,bal2024continuous}. We note that the absence of regularization, i.e. for $\eta=0$, the operator $H_I$ is not elliptic when $v_x(y)$ is allowed to vanish for at such points the singular values of $\alpha$ do not grow linearly as $\xi\to\infty$.
.

\section{Relation to dielectric tensor in photonics}

Let $\psi=(E,B)^T$ and recast the spectral problem for the cold plasma Hamiltonian as
\[ \begin{pmatrix} A & B \\ B^* & C\end{pmatrix}\begin{pmatrix} v \\ \psi \end{pmatrix}
=
\omega\begin{pmatrix} v \\ \psi \end{pmatrix}.
\]
Elimination of $v$ assuming $\omega$ not in the spectrum of $A$ yields by Schur complement the relation $(C- B^*(A-\omega) B)\psi = \omega \psi$ 
or equivalently 
\[ C \psi = \omega M(\omega) \psi,\qquad M(\omega) := I- \frac1\omega B^*(\omega-A)^{-1}B .\]
For the Hamiltonian in \eqref{eq:HFourier}, we find
\[  M= {\rm Diag} (\bar\epsilon,I_3),\qquad \bar\epsilon = 
\begin{pmatrix}
    \epsilon_1 & \epsilon_{12} & 0 \\ -\epsilon_{12} & \epsilon_1 & 0 \\ 0 & 0 & \epsilon_3
\end{pmatrix}\]
where the coefficients in the plasma dielectric tensor $\epsilon$ are given as in \cite{qin2023topological,silveirinha2015chern,hanson2016notes} by $\epsilon_1=1-\frac{\omega_p^2}{\omega^2-\Omega^2}$, $\epsilon_3=1-\frac{\omega_p^2}{\omega^2}$, and the non-reciprocal component $\epsilon_{12} = i \frac{-\Omega\omega_p^2}{\omega(\omega^2-\Omega^2)}$.

The analysis of topological phases presented in the paper presumably extends to more general models of photonic materials \cite{silveirinha2015chern,hanson2016notes,han2022anomalous} and this will be explored in future work.

Let us conclude this section with a comment on the computation of Berry connections $\mA$ in the context of the above elimination of the component $v$. Let us assume that $A$ and $B$ are independent of $\kp$, which holds in the absence of regularization. The connection of the full field $(v,\psi)^T$ is defined for the above model as:
\[
 v\cdot dv + \psi \cdot d\psi = (B^* (\omega-A)^{-2} B +I)\psi  \cdot d\psi = \psi\cdot \partial_\omega(\omega-B^*(\omega-A)^{-1}B) d\psi.
\]
Note that the normalization of eigenvectors is
\[
 v\cdot v + \psi \cdot \psi = (B^* (\omega-A)^{-2} B +I)\psi  \cdot \psi = \psi\cdot \partial_\omega(\omega-B^*(\omega-A)^{-1}B) \psi.
\]
Thus, 
\[
 v\cdot dv + \psi \cdot d\psi = \frac{v\cdot dv + \psi \cdot d\psi}{v\cdot v + \psi \cdot \psi} = \frac{\psi\cdot \partial_\omega(\omega M(\omega)) d\psi}{\psi\cdot \partial_\omega(\omega M(\omega)) \psi},
\]
which is the formula for the computation of Berry connections and Berry curvatures considered in \cite{PhysRevLett.100.013904,silveirinha2015chern} as the natural expression of Berry connections after elimination of the component $v$. We note that the above formula applies more generally for $M=M(\kp,\omega)$ \cite{PhysRevLett.100.013904,silveirinha2015chern}.

\section{Regularization}
\label{sec:regul}

This section provides some possible heuristic justification for the regularization proposed in \eqref{eq:regulOm}. We recast the electron transport equation for current $\tilde v$ as
\[ \partial_t \tilde v = \Omega e_z\times \tilde v -\omega_p \tilde E.\]
Let us now assume as in \cite{pakniyat2022chern,serra2025influence} the presence of a density term $\rho$ satisfying the continuity equation $\partial_t \rho + \nabla\cdot \tilde v=0$ and modifying the above transport equation to include collisions as
\[ \partial_t \tilde v +\beta^2 \nabla \rho  = \Omega e_z\times \tilde v -\omega_p \tilde E.\]
These equations in dual variables are given by (with the convention $i\partial_t\equiv\omega$)
\[  \omega\tilde v - \beta^2 \kop \rho = i\Omega e_z\times \tilde v - i \omega_p \tilde E,\qquad -\omega \rho + \kop\cdot \tilde v=0.\]
Eliminating $\rho$ gives
\[
\omega (1 - \frac{\beta^2 \kop\otimes \kop}{\omega^2}) \tilde v =i\Omega e_z\times \tilde v - i \omega_p \tilde E.
\]

Define
\[v= (1+\frac{\beta^2}{\omega^2} \kop\otimes \kop)^{\frac12} \tilde v, \quad  E= (1+\frac{\beta^2}{\omega^2} \kop\otimes \kop)^{-\frac12} \tilde E \]
to obtain
\[
  (1 - \frac{\beta^2 \kop\otimes \kop}{\omega^2})(1 + \frac{\beta^2 \kop\otimes \kop}{\omega^2})^{-1} \omega v = 
  i\Omega (1+\frac{\beta^2}{\omega^2} \kop\otimes \kop)^{-\frac12}e_z\times(1+\frac{\beta^2}{\omega^2} \kop\otimes \kop)^{-\frac12} v - i \omega_p E.
\]
Denoting by $\Pi=\hat\kop\otimes\hat\kop$ the projector onto the direction $\hat\kop=\kop/k$, we observe that the term on the left-hand side is
\[
\Big( I-\Pi + \frac{1-\frac{\beta^2k^2}{\omega^2}}{1+\frac{\beta^2k^2}{\omega^2}}\Pi\Big) \omega v.
\]
For $|\kop|$ large, we further obtain after some algebra that:
\[
 (1+\frac{\beta^2}{\omega^2} \kop\otimes \kop)^{-\frac12} \Omega e_z \times (1+\frac{\beta^2}{\omega^2} \kop\otimes \kop)^{-\frac12} \approx (1+\frac{\beta^2k^2}{\omega^2})^{-\frac12} \Omega e_z\times.
\]
Therefore, in that limit
\[ \omega (I-2\Pi)v \approx (1+\frac{\beta^2k^2}{\omega^2})^{-\frac12} \Omega e_z\times -i\omega_p E.\]
We observe that the component of $\Psi_{1,2}$ in \eqref{eq:psiinf} that needs regularization is $v_2$, i.e., the component orthogonal to $\kop=(k_x,0,k_z)$. This is therefore a component in the kernel of $\kop\otimes\kop$ for which the above with $\Pi v=0$ applies.  This provides some heuristics for the regularization introduced in \eqref{eq:regulOm} for any band where $\omega$ does not converge to $0$ as $k\to\infty$, which is the case for the bands we are interested in.  We stress again that the numerical simulations performed in section \ref{sec:main} were obtained without any regularization of the Hamiltonian. We do not have a complete justification of \eqref{eq:BEC} for either regularized or un-regularized Hamiltonians.
\section{Reduced two-band models}
\label{sec:Dirac}


This section provides additional information on the derivation of \eqref{eq:DiracM} and \eqref{eq:Dirac} when $\Omega\to-\Omega$. The degenerate eigenvectors associated to $\op = \omega_-$ when $\bk = 0, k_z, \om > 0$, denoted $\psi_{10}, \psi_{20}$ are:
\[  \psi_{10} = \frac{1}{\sqrt{2}}(\hat{e}_z, i\hat{e}_z, 0)^T,\quad  \psi_{20} = c\left(i\frac{k_z^2}{|\om|}\hat{e}_+, \omega_-\hat{e}_+, -ik_z \hat{e}_+\right)^T\]
for an appropriate normalizing constant $c$. To obtain a 2x2 approximation of $H(\bk)$ around the point $\bk = 0, \op = \omega_-$ we then project $H(\bk)$ onto $\psi_{10}, \psi_{20}$. Since the model is an approximation around $\bk = 0$ and $\op = \omega_-$ we split $H$ into:
\[
H(\bk) = H_0 + \tilde{H}(\bk) =
\begin{pmatrix}
    i\om \hat{e}_z \times & -i\omega_- & 0\\
    i\omega_- & 0 & -k_z \hat{e}_z \times\\
    0 & k_z \hat{e}_z \times & 0
\end{pmatrix}
+ \begin{pmatrix}
    0 & -i\tilde{\omega} & 0\\
    i\tilde{\omega} & 0 & -\bk \times\\
    0 & \bk \times & 0
\end{pmatrix}.
\]
When $\tilde{\omega}$ is a small positive value the $H(\bk)$ is in phase $\Phb^+$ and when $\tilde{\omega}$ is small an negative $H(\bk)$ is in phase $\Pha^+$. Clearly $\psi_{10}, \psi_{20}$ are eigenvectors of $H_0$ both with eigenvalue $\omega_-$. With $\hat z\times e_+=-ie_+$ we find
\[
\tilde{H}(\kp)\psi_{20} = c \begin{pmatrix} -i \tilde{\omega} \omega_- \hat{e}_+ \\ (-\frac{k_z^2}{|\om|} \tilde{\omega} +ik_z \bk \times )\hat e_+ \\ \omega_- \bk \times \hat e_+ \end{pmatrix}\
\bk \times \hat e_+ = \frac{1}{\sqrt2} \begin{pmatrix} 0 \\ 0\\ ik_x-k_y\end{pmatrix} 
\ \bk \times \hat e_- = \frac{1}{\sqrt2} \begin{pmatrix} 0 \\ 0\\ -ik_x-k_y\end{pmatrix},
\]
from which we deduce, since $(\hat e_z,\hat e_\pm)=0$:
\[
(\psi_{10},\tilde{H}(\kp)\psi_{20}) = \frac{ck_z}{\sqrt{2}} (ik_x-k_y),\quad(\psi_{20},\tilde{H}(\kp)\psi_{10}) = -\frac{k_z c}{\sqrt{2}} (ik_x + k_y),
\]
\[(\psi_{10},\tilde{H}(\bk)\psi_{10}) = \tilde{\omega}, \quad (\psi_{20},\tilde{H}(\bk)\psi_{20}) = \frac{-2\omega_-  c^2 k_z^2}{|\om|}\tilde{\omega}.
\]
After explicit calculation of $c$ this yields \eqref{eq:Dirac}. Now note that $S(\theta)\hat{e}_z=\hat{e}_z$ so that $\tilde\Gamma_P(\theta) \psi_{10}=\psi_{10}$, but $S(\theta)\hat{e}_+=e^{i 2\theta}\hat{e}_-$ implying that 
\[\tilde\Gamma_P(\theta) \psi_{20}=e^{i2\theta} (i\frac{k_z^2}{|\om|} \hat{e}_-,\omega_-\hat{e}_-,ik_z\hat{e}_-)^T.\]
This phase will necessarily cancel out in the diagonal elements $(\psi_{20},\tilde{H}(\kp)\psi_{20})$ and $(\psi_{10},\tilde{H}(\bk)\psi_{10})$ but for the off-diagonal terms we obtain:
\[
(\psi_{10}, H(\bk) \tilde \Gamma_P(\theta)\psi_{20})=   e^{i2\theta} \frac{k_z c}{\sqrt{2}} (ik_x+k_y), \quad (\psi_{20}, H(\bk) \tilde \Gamma_P(\theta)\psi_{10}) = e^{i2\theta} \frac{ck_z}{\sqrt{2}} (k_y-ik_x).
\]
Again with explicit calculation of $c$ we obtain \eqref{eq:DiracM}. $\theta$ is arbitrary since $\bk = 0$ for $\psi_{10}, \psi_{20}$. Therefore this shows that the transition from $\Omega$ to $-\Omega$ yields $(k_x,k_y)\to(k_x,-k_y)$ for $\theta=0$ while $(k_x,k_y)\to(-k_x,k_y)$ for $\theta=\frac\pi2$ for instance. This arbitrary phase could have also been deduced from adding an arbitrary phase to $\psi_{10}$ or $\psi_{20}$ and reflects the rotational invariance of the problem. This justifies the form of the models deduced in section \ref{sec:reduced}.

\section{Numerical simulations}
We follow a similar finite difference scheme as \cite{fu2021topological}\cite{bal2024topological} to numerically calculate the spectrum of the interface Hamiltonian $H_I$. We choose to vary the parameters $\op, \om$ in the $y$-direction so that the Hamiltonian is invariant with respect to translations in $x$. Taking the Fourier transform of \eqref{eq:HI} in $t, x, z$ gives:
\[
\omega \psi(y) = 
\begin{pmatrix}
    i \om(y)\hat{e}_z \times & -i\op(y) & 0\\
    i\op(y) & 0 &  i \partial_y\hat{e}_y \times -(k_x, 0, k_z)^t \times \\
    0 & -i \partial_y\hat{e}_y \times  + (k_x, 0, k_z)^t \times & 0
\end{pmatrix}
\psi(y)
\]
which is an ODE in $y$ for each $k_x$. Discretizing $y$ into $N$ points on an interval $[-L, L]$ allows us to approximate the equation as:
\begin{equation}\label{discH}
\omega\psi(y_i) =
H_i^-\psi(y_{i-1}) + 
H_i \psi(y_i)  + H_i^+ \psi(y_{i+1}) 
\end{equation}
\[
H_i = 
\begin{pmatrix}
    i \om(y_i)\hat{e}_z \times & -i\op(y_i) & 0\\
    i\op(y_i) & 0 &  \frac{i}{\Delta y} \hat{e}_y\times -\frac12(k_x, 0, k_z)^t \times \\
    0 & -\frac{i}{\Delta y}\hat{e}_y\times + \frac12(k_x, 0, k_z)^t \times & 0
\end{pmatrix}
\]
\[
H_i^- = 
\begin{pmatrix}
    0 & 0 & 0\\
    0 & 0 & -D\\
    0 & 0 & 0
\end{pmatrix}
\;\;\;\;\;\;
H_i^+ = 
\begin{pmatrix}
    0 & 0 & 0\\
    0 & 0 & 0\\
    0 & D & 0
\end{pmatrix} \;\; D = \frac{i}{\Delta y}\hat{e}_y\times + \frac12 (k_x, 0, k_z)^t\times
\]
where $\Delta y = 2L/N$. We adopt the particular combination of forward and backward differences and averaging $E, B$ in the second and third row from \cite{fu2021topological} to ensure the discrete problem obeys the same particle-hole symmetry as the continuous one. This convention amounts to discretizing $B$ on half-integer grid points. Calculating the eigenvalues of $H_I$ can then be done by diagonalizing a $9N \times 9N$ matrix. In order to ensure the matrix is Hermitian periodic boundary conditions are enforced ($\psi(-L) = \psi(L)$, $\op(-L) = \op(L)$, $\om(-L) = \om(L)$). Assuming our parameters cross from one topological phase into another, having periodic parameters $\om, \op$ introduces a second topological transition in the opposite direction, which will necessarily produce edge states concentrated around this second edge (in cases where edge states exist). For clarity, we adopt the strategy used in \cite{bal2024topological} and eliminate eigenvalues whose associated eigenvector has more than half its weight within $0.1L$ of the spurious edge. 

\begin{figure}[ht]
			\centering
			\includegraphics[width=0.9\textwidth]{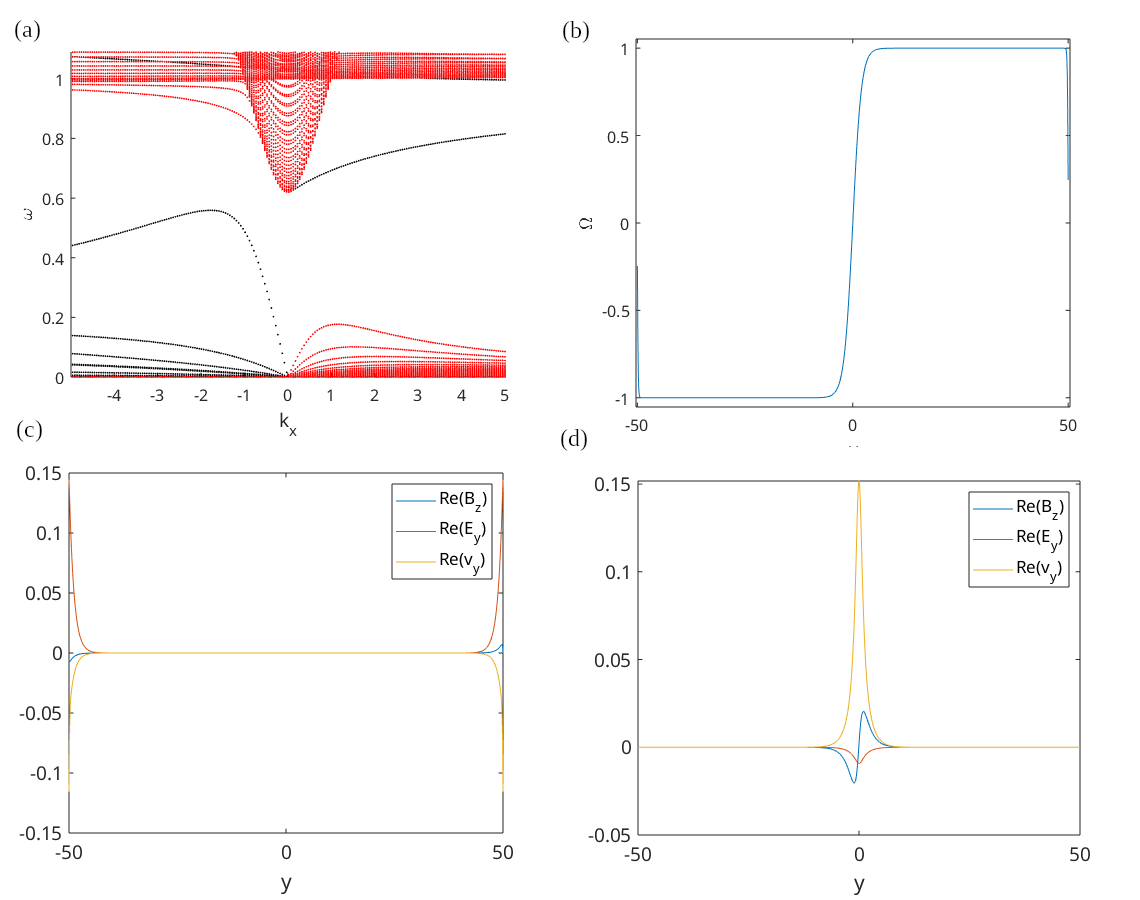}
		\caption{Illustration of the elimination of spurious edge modes for the spectrum calculated in Figure 6. (a) shows the spectrum with spurious modes shown in black and bulk and legitimate edge modes in red. (b) is the profile of $\om$, which consists of a periodic combination of logistic functions. (c) and (d) plot components of the numerically calculated eigenvectors plotted as a function of $y$. (c) is a mode concentrated around the spurious edge, specifically the eigenvector associated to $(k_x, \omega) = (1, 0.6942)$, while (d) is a legitimate edge mode at ($(k_x, \omega) = (-1, 0.8724)$). Modes of the former type are eliminated for clarity as discussed above.}
        \label{fig:numerics_1}
\end{figure}

\begin{figure}[ht]
			\centering
			\includegraphics[width=0.5\textwidth]{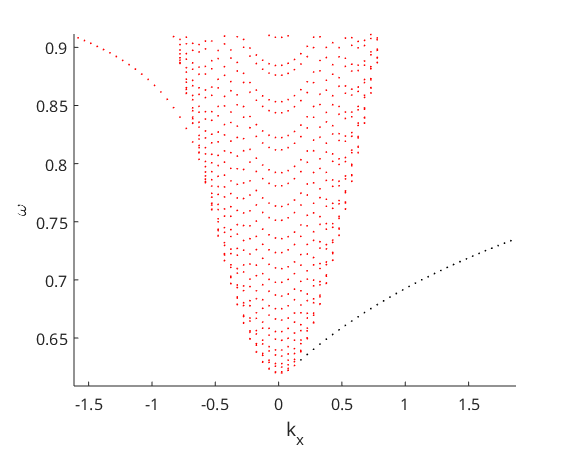}
		\caption{Inset of Figure \ref{fig:numerics_1} which illustrates one branch described in the last paragraph of Appendix E which exhibit both spurious and edge modes, lowest branch which can be seen transitioning from red to black at a small positive value of $k_x$.}
        \label{fig:numerics_2}
\end{figure}

One disadvantage of this method is the appearance of continuous branches of spectrum which are edge modes concentrated at $y = 0$ for $k_x <0$ ($k_x >0$) and concentrated at $y = L$ for $k_x > 0$ ($k_x <0$). Figure 8 illustrates one such branch. Since we have two topological phase transitions in opposite directions, we expect that each uni-directional edge state localized at $y = 0$ for $k_{x0} <0$ will be mirrored by a uni-directional edge state localized at $y = L$ for $k_{xL} = -k_{x0}$. For bands which cross a spectral gap (topologically protected edge states) this produces a mirrored band localized at $y = L$ which produces an identical spectral flow in the opposite direction (see \cite{fu2021topological, qin2023topological}). However for edge states whose energy does not overlap with the spectral gap, bands may appear to switch from concentrated at $y = 0$ for $k_x <0$ ($k_x >0$) to concentrated at $y = L$ for $k_x > 0$ ($k_x <0$) as we see in Figure 9. The appearance of these bands is therefore due to the periodic boundary conditions we impose and we expect that such bands would not appear when considering more realistic boundary conditions. Inclusion of the spurious modes shows as in Figures 8 and 9 shows that these branches indeed only exist at energies outside the spectral gap and therefore do not contribute to $\sigma_I$. 

\section{Topological Phase and Explicit Eigenvector Calculations}
First we prove the inequality $\omega_{L^-} < \omega_R < \omega_{L^+}$. This can be proved from \eqref{cubicpm}, copied here for in a more convenient form for this derivation:
\[
    k_z^2 = \omega_R^2 - \frac{\omega_p^2}{1+\frac{\Omega}{\omega_R}},\qquad k_z^2 = \omega_L^2 - \frac{\omega_p^2}{1-\frac{\Omega}{\omega_L}}.
\]
Notice from \eqref{cubicpm} that as $k_z \rightarrow \infty$,   $\omega_L$ has two positive solutions, $\omega_L \rightarrow \om$ from below, and $\omega_L \rightarrow \sqrt{k_z^2 + \op^2} \rightarrow \infty$. By continuity of eigenvalues then we know that there are two positive branches of $\omega_L$, one of which is $0 < \omega_{L^-} < \Omega$ and one of which is $\Omega < \omega_{L^+}$ since $k_z^2 \rightarrow \pm \infty$ if $\omega_L$ approaches $\om$, which would violate continuity of $\omega_{L^\pm}$ as a function of $k_z$. 
Also notice that for $\om> 0$ and $\omega_R > 0$:
\[
0 < \frac{\omega_p^2}{1+\frac{\om}{\omega_R}} < \omega_p^2 \ \ \Rightarrow  \ \ 
k_z^2 < \omega_R^2 < \omega_p^2 + k_z^2.
\]
Next since we know that $\omega_{L^+} > \Omega$ so that we also have $0 < \om/\omega_{L^+} < 1$ so that:
\[
\omega_{L^+}^2 = k_z^2 + \frac{\omega_p^2}{1-\frac{\om}{\omega_{L^+}}} > k_z^2 + \op^2
\]
so indeed $\omega_{L^+} > \omega_R$ and in addition $\omega_{L^+} > \omega_p$. Finally since $\omega_{L^-} < \Omega \Rightarrow \om/\omega_{L^-} > 1$:
\[
\omega_{L^-}^2-k_z^2 = \frac{\omega_p^2}{1-\frac{\om}{\omega_{L^-}}} < 0 \ \ \Rightarrow  \ \ 
\omega_{L^-}^2 < k_z^2 .
\]
To summarize we have proved that:
\[
\omega_{L^-}^2 < \min\{\om^2, k_z^2\} \le k_z^2 <\omega_R < k_z^2 + \op^2 < \omega_{L^+},
\]
so indeed $\omega_{L^-} < \omega_R < \omega_{L^+}$. This proves that when $k = 0$ the only band crossings occur at $\op = \omega_\pm$. Again we rely on extensive numerical evidence that band crossings do not occur except when $k = 0$, $k_z = 0$, or $\om = 0$. Therefore the topological phases are completely determined by the curves $k_z = 0$, $\om  =0$, and $\omega_p = \omega_\pm$, which divide the parameter space into phases $\Pha^\pm, \Phb^\pm, \Phc^\pm, \Phd^\pm$ as shown in Figure 1.

We also wish to point out more explicit forms of \eqref{eq:psiinf}, \eqref{eq:symH}, \eqref{eq:invOmega} which make the gluing conditions more explicit. \eqref{eq:psiinf} represents the eigenvectors as $k \rightarrow \infty$ for $\mathbf{k} = k \hat{e}_x$. Using  \eqref{eq:invrot} we can obtain eigenvectors for arbitrary $\mathbf{k}$ by applying R$(\theta)$ to $\Psi_n$, which amounts to applying $R(\theta)$ to each of $v, E, B$ components of $\Psi_n$. Denote the unit $\mathbf{k}$ direction $\hat{\mathbf{k}} = \mathbf{k}/|\mathbf{k}|$. Noting that $R(\theta)$ leaves the $\hat{e}_z$ component invariant and $\hat{e}_y = \hat{e}_z \times \hat{e}_x = \hat{e}_z \times \hat{\mathbf{k}}(\theta = 0)$ we get:
\[
\Psi_1(\bk, \om) = 
\begin{pmatrix}
    -\frac{1}{\sqrt{1 + \sigma^2}}R(\theta)(\hat{e}_z \times \hat{\bk}(\theta = 0)) + i\hat{e}_z\\
    -\frac{\sigma}{\sqrt{1+\sigma^2}} R(\theta) \hat{\bk}(\theta = 0)\\
    0
\end{pmatrix}
= \begin{pmatrix}
    \frac{1}{\sqrt{1+\sigma^2}}\hat{\bk} \times \hat{e}_z + i\hat{e}_z\\
    -\frac{\sigma}{\sqrt{1+\sigma^2}}\hat{\bk}\\
    0
\end{pmatrix}.
\]
Similarly for the remaining eigenvectors as $k\rightarrow \infty$:
\[
\Psi_2(\bk, \om) = \begin{pmatrix}
    \frac{\sigma}{\sqrt{1+\sigma^2}}\hat{\bk} \times \hat{e}_z - i\hat\bk\\
    \frac{1}{\sqrt{1+\sigma^2}} \hat{\bk}\\
    0
\end{pmatrix}
\;\;\;
\Psi_3(\bk, \om) = 
\begin{pmatrix}
    0\\
    \hat{e}_z\\
    \hat{\bk} \times \hat{e}_z
\end{pmatrix}
\;\;\;
\Psi_4(\bk, \om) = 
\begin{pmatrix}
    0\\
    -\hat{\bk} \times \hat{e}_z\\
    \hat{e}_z
\end{pmatrix}
\]
where these expressions are valid assuming $\om >0$. 

As mentioned in Section 3.2 the Hamiltonian also obeys parity symmetry under the anti-unitary operator $\Gamma_k K$, where $K$ is the operator performing element-wise complex conjugation such that $\Gamma_k K H(\theta) K\Gamma_k = -H(\theta)$. To be more explicit we have $H(\bk) (\Gamma_k K\Psi_n(\bk)) = -\omega_n(\Gamma_k K\Psi_n(\bk))= \omega_{-n}(\Gamma_k K\Psi_n(\bk))$. Therefore the corresponding eigenvectors for $\omega \rightarrow -\omega$ are:
\[
\Psi_{-1}(\bk, \om) = 
\begin{pmatrix}
    \frac{1}{\sqrt{1+\sigma^2}}\hat{\bk} \times \hat{e}_z - i\hat{e}_z\\
    -\frac{\sigma}{\sqrt{1+\sigma^2}}\hat{\bk}\\
    0
\end{pmatrix}\;\;\;
\Psi_{-2}(\bk, \om) = \begin{pmatrix}
    \frac{\sigma}{\sqrt{1+\sigma^2}}\hat{\bk} \times \hat{e}_z + i\hat\bk\\
    \frac{1}{\sqrt{1+\sigma^2}} \hat{\bk}\\
    0
\end{pmatrix}
\;\;\;
\]
\[
\Psi_{-3}(\bk, \om) = 
\begin{pmatrix}
    0\\
    \hat{e}_z\\
    -\hat{\bk} \times \hat{e}_z
\end{pmatrix}
\;\;\;
\Psi_{-4}(\bk, \om) = 
\begin{pmatrix}
    0\\
    -\hat{\bk} \times \hat{e}_z\\
    -\hat{e}_z
\end{pmatrix}.
\]
Then by combining with $\Gamma_\om$ from Section 3.2 ($\Gamma_\om H(\bk, \om)\Gamma_\om = -H(\bk, -\om)$), we get $H(\bk, -\om) = \Gamma_k K \Gamma_\om H(\bk, \om)\Gamma_\om K\Gamma_k $ and subsequently:
\[
\Psi_{1}(\bk, -\om) = 
\begin{pmatrix}
    \frac{1}{\sqrt{1+\sigma^2}}\hat{\bk} \times \hat{e}_z - i\hat{e}_z\\
    \frac{\sigma}{\sqrt{1+\sigma^2}}\hat{\bk}\\
    0
\end{pmatrix}\;\;\;
\Psi_{2}(\bk, -\om) = \begin{pmatrix}
    \frac{\sigma}{\sqrt{1+\sigma^2}}\hat{\bk} \times \hat{e}_z + i\hat\bk\\
    -\frac{1}{\sqrt{1+\sigma^2}} \hat{\bk}\\
    0
\end{pmatrix}
\;\;\;
\]
\[
\Psi_{3}(\bk, -\om) = 
\begin{pmatrix}
    0\\
    -\hat{e}_z\\
    -\hat{\bk} \times \hat{e}_z
\end{pmatrix}
\;\;\;
\Psi_{4}(\bk, -\om) = 
\begin{pmatrix}
    0\\
    \hat{\bk} \times \hat{e}_z\\
    -\hat{e}_z
\end{pmatrix}
\]
\[
\Psi_{-1}(\bk, -\om) = 
\begin{pmatrix}
    \frac{1}{\sqrt{1+\sigma^2}}\hat{\bk} \times \hat{e}_z + i\hat{e}_z\\
    \frac{\sigma}{\sqrt{1+\sigma^2}}\hat{\bk}\\
    0
\end{pmatrix}\;\;\;
\Psi_{-2}(\bk, -\om) = \begin{pmatrix}
    \frac{\sigma}{\sqrt{1+\sigma^2}}\hat{\bk} \times \hat{e}_z - i\hat\bk\\
    -\frac{1}{\sqrt{1+\sigma^2}} \hat{\bk}\\
    0
\end{pmatrix}
\;\;\;
\]
\[
\Psi_{-3}(\bk, -\om) = 
\begin{pmatrix}
    0\\
    -\hat{e}_z\\
    \hat{\bk} \times \hat{e}_z
\end{pmatrix}
\;\;\;
\Psi_{-4}(\bk, -\om) = 
\begin{pmatrix}
    0\\
    \hat{\bk} \times \hat{e}_z\\
    \hat{e}_z
\end{pmatrix}.
\]
This is an exhaustive list of limiting eigenvectors as $k \rightarrow \infty$ in any case except the singular cases $k_z = 0$ and $\om = 0$. It is clear that as long as $\sigma(k) \rightarrow \bar{\sigma}$ independent of topological phase that $\Pi_n^N$ glues to $\Pi_n^S$ as long as $\om$ has the same sign for both $N$ and $S$. However as soon as $\om$ changes sign this is not the case. There are two regularizations which produce integers for $\mC[P_\ell^h]$ and hence integers for $\fC[P_\ell^S, P_\ell^N]$ when $\om$ changes sign, namely $\op(k)\rightarrow 0$ and $\om \rightarrow 0$ respectively as $k\rightarrow \infty$, which correspond to $\sigma \rightarrow \infty$ and $\sigma \rightarrow 0$ respectively. Comparing eigenvectors in these two limits we see that in the $\op(k)\rightarrow 0$ case:
\[
   \Pi_j(\bk,-\om) = \Pi_j(\bk,\om),\ j\in\{1,3,4\};\quad 
   \Pi_2(\bk, -\om) = \Pi_{-2}(\bk, \om).
\]
So while \eqref{eq:gluing} holds for $P_\ell^{N/S}$ when $\ell = 2$ regardless of a change of sign in $\om$, it does not hold for $\ell = 1$. Meanwhile for $\om \rightarrow 0$ we get:
\[
   \Pi_1(\bk, -\om) = \Pi_{-1}(\bk, \om); \quad 
   \Pi_j(\bk, -\om) = \Pi_j(\bk, \om) ,\ \ j\in\{2,3,4\},
\]
so \eqref{eq:gluing} holds for $\ell = 1, 2$ regardless of a change of sign of $\om$.

\bibliographystyle{unsrt}
\bibliography{Refs}

\end{document}